\documentclass[pre,a4paper,onecolumn,notitlepage,nofootinbib]{revtex4-1}
\usepackage{graphicx}
\usepackage{amsmath}
\usepackage{bbm}
\usepackage{psfrag}
\usepackage{latexsym}

\def\be{\begin{eqnarray}}
\def\ee{\end{eqnarray}}

\newcommand{\Sg}{\Sigma}

\newcommand{\eq}{\begin{equation}}
\newcommand{\eqx}{\end{equation}}
\newcommand{\eqn}{\begin{eqnarray}}
\newcommand{\eqnx}{\end{eqnarray}}
\newcommand{\ben}{\begin{eqnaray}}
\newcommand{\een}{\end{eqnarray}}
\newcommand{\f}[2]{\frac{#1}{#2}}

\newcommand{\GG}{{\cal G}}

\newcommand{\ZZ}{{\cal Z}}
\renewcommand{\SS}{{\cal S}}
\newcommand{\WW}{{\cal W}}
\newcommand{\YY}{{\cal Y}}
\newcommand{\DD}{{\cal D}}
\newcommand{\HH}{{\cal H}}
\newcommand{\RR}{{\cal R}}

\newcommand{\zb}{\bar{z}}

\newcommand{\arr}[4]{
\left(\begin{array}{cc}
#1&#2\\
#3&#4
\end{array}\right)
}

\newcommand{\cor}[1]{\left\langle{#1}\right\rangle}

\newcommand{\tr}{\mbox{\rm tr}\,}

\newcommand{\lm}{\lambda}

\newcommand{\sg}{\sigma}

\newcommand{\oo}[1]{{\mathcal{O}}\left(#1\right)}

\begin{document}

\title{\bf Multiplication law and S transform for non-hermitian 
random matrices}

\author{Z. Burda}\email{zdzislaw.burda@uj.edu.pl}

\author{R.A. Janik}\email{romuald@th.if.uj.edu.pl}

\author{M.A. Nowak}\email{nowak@th.if.uj.edu.pl}

\affiliation{Marian Smoluchowski Institute of Physics and 
Mark Kac Complex Systems Research Center \\
Jagiellonian University, Reymonta 4, 30-059 Krak\'ow, Poland}

\date{\today}

\begin{abstract}
We derive a multiplication law for free non-hermitian random matrices 
allowing for an easy reconstruction of the two-dimensional eigenvalue distribution of the product ensemble from the characteristics of the individual
ensembles. We define the corresponding non-hermitian S transform being a natural generalization of the Voiculescu S transform. In addition we extend the classical hermitian S transform approach to deal with the situation when the random matrix ensemble factors have vanishing mean including the case when both of them are centered. We use planar diagrammatic techniques to derive these results.  

\medskip

\end{abstract}

\maketitle

\section{Introduction}
Free random variables~\cite{VOICULESCU,SPEICHER} play an increasingly 
important role in mathematics, physics, multivariate statistics and interdisciplinary research~\cite{TSE,TELATAR,MULLER,VERDU,BURDANOWAK,BOUCHAUD,QCD,RYAN}. 
The cornerstones of this success are the so-called R and S
transforms. The R transform allows one to infer the spectral
properties of the sum of random operators, provided the individual
spectral measures are known for each of them and they are independent
in the noncommutative sense also known as free. The S transform plays
a similar role for the multiplication of free random operators.
These constructions allow for fast decomposition of several problems
for complicated random operators into simple ingredients. 
Since free random operators have
an explicit realization in terms of infinitely large random matrices,
the techniques based on the R and S transforms provide a powerful tool 
to solve technically involved problems in random matrix theory in an easy way when
traditional methods break down. 

Historically, the R transform was devised
for hermitian operators and the S transform for unitary ones. 
The issue of the generalization of these constructions to
other classes of operators was a subject of intensive research during the last
two decades.  In particular, one of the most challenging problems
was the question of the possibility of an extension of the R and S
transforms to strictly non-hermitian matrices, which find nowadays vast 
applications in many fields of research.
This problem is also especially interesting as traditional techniques developed for hermitian random matrices generally fail in the non-hermitian case. Some
time ago, two of the present authors have extended the \emph{additive} 
R transform for the non-hermitian ensembles~\cite{JANNOWPAPZAHWAM,JANNOWPAPZAH}. Similar constructions were also proposed independently in~\cite{FEIZEE,CHAWAN}, and were soon generalized~\cite{JARNOW,ROD}. 
The question
of defining the \emph{multiplicative} S transform for non-hermitian matrices was however open 
and frequently doubts were expressed whether such a construction is possible at
all.  On the other hand several complicated problems involving products of large matrices have been solved using other methods and results were sometimes surprisingly simple~\cite{PZJ,SPERAJ,EGNJANJURNOW,GIRKOVLAD,BUR,LIVAN}, hinting at the possibility of a hidden mathematical  structure.   

 In  this work we demonstrate that such a structure -- the non-hermitian S transform --  exists and can be used as a powerful algorithm for solving the spectral problems of various products of random matrices. As a byproduct we also generalize the
ordinary `hermitian' multiplicative technique to matrix ensembles with
vanishing mean which was never done before.  

In Section \ref{Sec_main} we outline main results of the paper. In particular
we give the multiplication law for free non-hermitian matrices.

In the next two sections, in order to make the paper self-contained, we introduce diagrammatic techniques which will be the main tool for deriving the key results of this paper.

In Section \ref{Sec_hermit} we very briefly recall the formalism
to calculate the eigenvalue densities of large random hermitian matrices 
in the limit of matrix dimensions $N \rightarrow \infty$. We recall the
connection to planar diagrams and use the diagrammatic technique to 
give a simple proof of the addition law. 

In Section \ref{Sec_non_hermit} we repeat the discuss for non-hermitian
matrices. We show that the Green's function and the R transform are given by $2\times 2$ matrices and recall the formalism to handle this case. 

In Section \ref{Sec_multip}, which is the main section of this paper,
we first rederive the multiplication law for hermitian matrices using diagrammatic arguments and then we generalize the construction to non-hermitian matrices. We discuss the S transform
for this case and show that similarly to the nonhermitian versions of 
the R transform and the Green's function
it has a form of a $2\times 2$ matrix.

Finally in Section \ref{Sec_examples} we give examples of 
application of this law to practical calculations of the eigenvalue
density for a product of free matrices.
We conclude the paper with a short summary.

\section{Main results \label{Sec_main}}

In this section we shortly summarize the main results of this paper.
The key quantity of interest in random matrix theory is the eigenvalue density,
which may be equivalently expressed through the Green's function. The R and S
transforms satisfy functional relations with the Green's function and hence their knowledge is equivalent (in the hermitian case) to the knowledge of the eigenvalue density (or more precisely of its moments).

Explicitly, the standard form of the multiplication law of free large hermitian
matrices is given in terms of the S transform~\cite{VOICULESCU} just through
an ordinary product
\begin{equation}
S_{AB}(z) = S_A(z) S_B(z)
\label{S_SS}
\end{equation}
The S transform is a complex function of a complex variable 
and it is related to the R transform as follows
\be
S(z) = \frac{1}{R\left(z S(z)\right)}
\ee
The two relations given above hold only if matrices $A$ and $B$ 
are not centered: $\langle {\rm tr} A \rangle \ne 0$ and
$\langle {\rm tr} B \rangle \ne 0$. This means
the corresponding R transforms may not vanish at the origin 
of the complex plane $R_A(z=0) \ne 0$ and $R_B(z=0) \ne 0$. 
If either $R_A(0) = 0$ or $R_B(0)=0$ but not both, the corresponding 
S transforms do not exist, but one can still save the multiplication law~\cite{SPERAJ}.  The prescription~\cite{SPERAJ} breaks down when both means ( i.e. for $A$ and $B$) ensemble vanish. One of our main new results is that one can still write
down a multiplication law in terms of the R transform
in that case too, using the  the following set of equations
\be
R_{AB}(z) \!&\!=\!&\! R_A(w) R_B(v)  \nonumber \\
v \!&\!=\!&\! z R_A(w) 
\label{R_RR} \\ 
w \!&\!=\!&\! z R_B(v) \nonumber
\ee
which involves three complex variables $z,w,v$.
One can  eliminate $w$ and $v$ for given $R_A$ and $R_B$ to
obtain $R_{AB}(z)$. This set is equivalent to the standard equation
(\ref{S_SS}) when the matrices $A$ and $B$  are not centered but it is
also valid when either of the two matrices, or even both, are centered, making
this a more general formulation. This set of equations is quite handy in practical calculations too. One can use it to directly calculate
the R transform of the free product avoiding the determination of all 
auxiliary functions and the S transform in particular. 
Another advantage of these equations is that they can be generalized in a natural  manner to the case of free
multiplication of non-hermitian operators and thus they can be used
to determine the eigenvalue distribution of products of non-hermitian matrices
taken from independent random ensembles in the large $N$ limit. 

Before we write down the corresponding set of equations let us first recall 
that the Green's function for non-hermitian matrices are conveniently expressed
as two-by-two matrices with complex elements~\cite{JANNOWPAPZAHWAM,JANNOWPAPZAH}. This will be in detail explained in the paper. The R transform in this case is a map of a space of two-by-two complex matrices onto a space of two-by-two complex matrices
${\cal G} \rightarrow {\cal R}({\cal G})$. 
In order to distinguish this situation from the hermitian case (\ref{R_RR})
where functions and their arguments were complex numbers we shall 
use calligraphic letters to denote the corresponding two-by-two complex matrices. The law of free multiplication for non-hermitian matrices reads 
\be
\nonumber
\RR_{AB}(\GG) \!&\!=\!&\! 
\left[\RR_A(\GG_B)\right]^L \cdot \left[\RR_B(\GG_A)\right]^R \\
\left[\GG_{A}\right]^R \!&\!=\!&\! \GG \cdot \left[\RR_A(\GG_B)\right]^L 
\label{AB_G} \\
\left[\GG_{B}\right]^L \!&\!=\!&\! \left[\RR_B(\GG_A)\right]^R \cdot \GG  \, .
\nonumber
\ee
It has almost an identical algebraic structure as (\ref{R_RR}) except
that now all objects are two-by-two matrices and thus the order of
multiplications matters. The superscripts 
R and L outside the square brackets, which were absent in (\ref{R_RR}),
stand for right or left rotations, respectively, of a matrix 
$X$ in the brackets: $[X]^L = U X U^\dagger$ and $[X]^R = U^\dagger X U$.
The matrix $U$ is a unitary diagonal matrix 
$U={\rm diag} (e^{i\phi/4},e^{-i\phi/4})$ that depends on the 
phase $\phi$ of the complex number $z=|z|e^{i\phi}$ 
being the argument of the Green function ${\cal G} = {\cal G}(z,\bar{z})$ containing the information on the spectral distribution of complex eigenvalues on the complex plane $z$. 
Although this set of equations is more complicated than for hermitian matrices (\ref{R_RR}) it also gives a direct, practical way of determining the Green's function ${\cal G}$ of the product of random matrices $A$ and $B$. 
We will illustrate this by an explicit examples towards the end of the paper.
We will also introduce the S transform for non-hermitian matrices and
use it to rewrite the set of equations (\ref{AB_G}), however we think that from the operational point of view 
equations~(\ref{AB_G}) are more convenient.

\section{Hermitian matrices \label{Sec_hermit}}

\subsection{Preliminaries}
We are interested in finding the distribution of eigenvalues
$\lambda_i $, in the limit when  $N$ (the size of the matrix $H$) is
infinite. The average spectral distribution reads
 \be
\rho(\lambda)= \lim_{N\rightarrow \infty} 
\frac{1}{N} \left\langle \sum _{i=1}^N \delta
(\lambda - \lambda_i) \right\rangle \label{spect-h} 
\ee 
where $\lambda_i$ are eigenvalues of a random hermitian matrix $H$
and brakets $\cor{\ldots}$ denote averaging  over a given ensemble of $N \times
N$ random hermitian matrices generated with the probability 
\be P(H)
\propto e^{-N {\rm Tr} V(H)} . \label{probab} 
\ee 
For hermitian matrices eigenvalues $\lambda_i$'s lie on the real axis. 
It is convenient to introduce a complex-valued resolvent (Green's function) 
\be G(z)= \lim_{N\rightarrow \infty} 
\frac{1}{N} \left\langle {\rm Tr}\, \frac{1}{z \mathbbm{1}-H}\right\rangle \,. \label{green} 
\ee 
from which one can reconstruct the spectral density function (\ref{spect-h})
\be
\rho(\lambda) = \frac{1}{2\pi i} \,\lim_{\epsilon \rightarrow 0^+} \big(
G(\lambda -i\epsilon)-G(\lambda+i\epsilon) \big) \, . 
\label{recon} 
\ee
using the well-known formula $\frac{1}{\lambda \pm i
0^+} = {\rm  P.V.} \frac{1}{\lambda} \mp i\pi \delta(\lambda)$. 
The symbol $\mathbbm{1}$ will be used throughout the paper to denote
identity matrices of different size. Here it was an $N$-by-$N$ identity matrix.
The Green's function is a generating function for spectral moments $\mu_n=\lim_{N\rightarrow \infty} \frac{1}{N} \left \langle {\rm Tr} H^n \right \rangle$
\be G(z) 
= \sum_{n=0}^\infty \frac{\mu_n}{z^{n+1}}
\label{e.series} 
\ee 
with $\mu_0=1$, as follows from the $1/z$-expansion of (\ref{green}).
Another fundamental quantity is the "self-energy" $\Sigma=\Sigma(z)$ defined as 
\be 
G(z)=\frac{1}{ z-\Sigma(z)} \ .
\label{selfenergy} 
\ee
It is related to the Green's function by an independent equation
\be
\Sigma(z)=R(G(z)) \ , 
\label{SigmaR}
\ee
where the function 
\be
R(z)= 
\sum_{n=1}^{\infty} \kappa_n z^{n-1}
\label{Rdef}
\ee
is the generating function for planar connected moments 
$\kappa_n = \lim_{N\rightarrow \infty} \frac{1}{N} \langle\langle{\rm Tr} H^n\rangle\rangle$ called free cumulants and denoted by double brackets. 
This function is usually referred to as the  R transform.
Its form can be deduced from the integration measure~(\ref{probab}). 
The difference between the planar connected moments (free cumulants) 
$\kappa_n$ in (\ref{Rdef}) and the spectral moments $\mu_n$
(\ref{e.series}) will be explained in the next section where 
a diagrammatic interpretation of these equations will be discussed.

The relation between the generating function for spectral moments 
$G(z)$ and the generating function for connected moments $R(z)$
can be made explicit  if one eliminates $\Sigma$ from (\ref{selfenergy})
and (\ref{SigmaR}). This yields a relation 
\be
G(z) = \frac{1}{z-R(G(z))}
\label{GfromR}
\ee
which is equivalent to 
\be
G\left(R(z)+\frac{1}{z}\right)=z \, .
\label{Rtransform} 
\ee
One can use these relations to determine $G(z)$ for given $R(z)$ 
or vice versa. To give an example, consider the simplest case of a random
matrix from the Gaussian Unitary Ensemble (GUE). In this case
the only non-vanishing cumulant is $\kappa_2$. Without loss of generality
we can choose $\kappa_2=1$, so that $R(z)=z$. Using (\ref{GfromR})
we have $G(z) = 1/(z-G(z))$. The last equation
can be easily solved for $G(z)$ and the solution can be used to 
calculate the spectral density (\ref{recon}). One recovers the Wigner's semicircle $\rho(\lambda)=\frac{1}{2\pi}\sqrt{4-\lambda^2}$~\cite{Wigsemi}.

\subsection{Planar diagrams}

\begin{figure}
\begin{center}
\includegraphics[width=7cm]{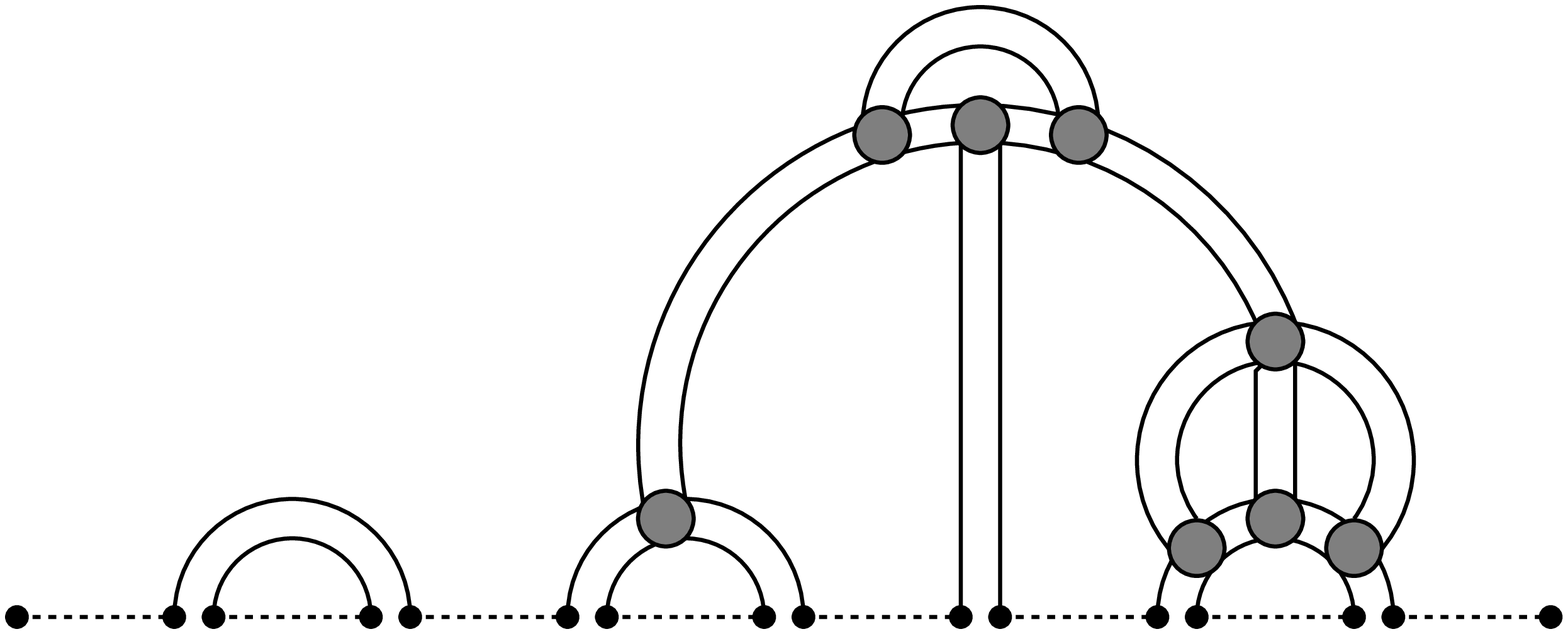} \qquad
\includegraphics[width=5cm]{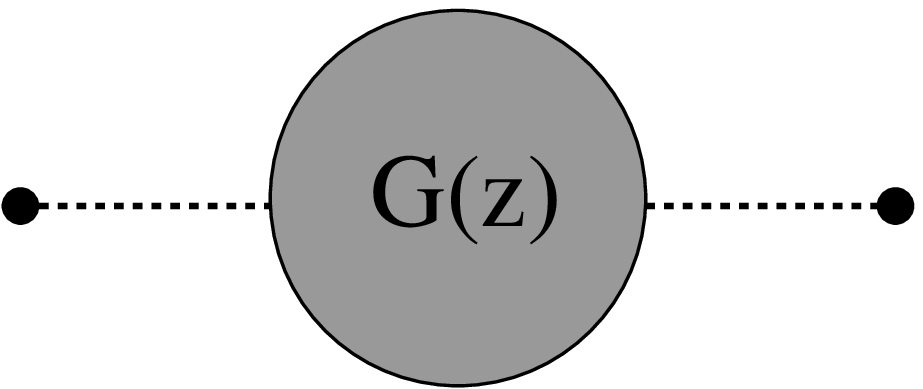}
\end{center}
\caption{(Left) An example of a diagram contributing to the 
generating function $G_{ij}(z)$. Two end-points 
should be labeled by indices $ij$. Each horizontal dashed line corresponds to $\frac{1}{z} \delta_{ab}$ while a double line represents the expectation value (the propagator) 
$\langle H_{ab} H_{cd} \rangle_0 = \frac{1}{N g_2} \delta_{ad} \delta_{bc}$.
Since all lines in the diagram are proportional to the 
delta function this equation reduces to a scalar equation for 
$G(z)$. Each horizontal dashed line corresponds after this reduction to $\frac{1}z$,
each double line to $\frac1{g_2}$, each vertex to $g_n$. The shown
diagram contributes
to the seventh moment $\frac{1}{N} \langle {\rm Tr} H^7\rangle$ which of order $\frac{1}{z^8}$ in the series expansion (\ref{e.series}) since it has eight
horizontal lines. The diagram contains seven cubic vertices $g_3^7$ and
one quartic vertex $g_4$ that are generated by the perturbative expansion
of the residual part of (\ref{split}).  
Each pair of dots on the horizontal line corresponds to a factor $H_{ab}$  
inside the average $\langle {\rm Tr} H^7 \rangle_0$. (Right) The graphical notation for the generating function $G(z)$. It generates diagrams having
two end-points which include for example the one shown on the left.
\label{Gexample}}
\end{figure}

One can calculate (\ref{e.series}) by Gaussian perturbation theory. 
One does it by splitting the integration
measure (\ref{probab}) into a Gaussian part and a 
residual part
\be
P(H) = {\cal N}^{-1}
e^{-N \frac{g_2}2 {\rm Tr} H^2} e^{- N \sum_{n \ne 2} \frac{g_n}n {\rm Tr} H^n}
= P_0(H) e^{- N \sum_{n\ne 2} \frac{g_n}n {\rm Tr} H^n} 
\label{split}
\ee
The Gaussian part $P_0(H)$ is then used to calculate averages 
$\langle \ldots \rangle_0$ while the remaining expression is left
inside the brackets and is averaged with respect to $P_0$. 
The constant ${\cal N}$ is an overall normalization.
This non-Gaussian  part is perturbatively expanded in $g_n$, so effectively one
has to calculate averages of various powers of $H$ with respect to 
the Gaussian measure. Each term in this expansion has a graphical 
representation, similar  to  Feynman diagrams known from quantum field theory (see figure \ref{Gexample}).
For example, single  horizontal lines  represent contributions
from the factors $\frac{1}{z} \mathbbm{1}$ in (\ref{e.series}).
In the large $N$ limit only planar diagrams contribute to  $G(z)$,
since all others are suppressed by $O(1/N)$ factors (note that each closed line generates a factor $N$ coming from  contraction of indices $\delta_{ii}=N$). 
Thus the calculation of $G(z)$ amounts to summing all (infinitely many) contributions from planar diagrams with 
two endpoints as shown in figure (\ref{Gexample}). 
Actually in the most general case one should rather consider a
matrix form of the Green's function $\mathbbm{G} = (G_{ij}(z))$
where $i$ and $j$ are indices of two end-points $i=1,\ldots,N$, $j=1,\ldots,N$ 
(see figure \ref{Gexample}) and calculate the scalar function (\ref{green})
afterwards as the normalized trace $G(z) = \frac{1}{N} {\rm Tr} {\mathbbm G}(z)$. Also the self-energy equation (\ref{selfenergy}) should formally be written in a matrix form. However in our case all generating matrices are proportional to Kronecker delta functions
$z_{ij} = z \delta_{ij}$, $G_{ij}(z) = G(z) \delta_{ij}$, 
$\Sigma_{ij}(z) = \Sigma(z) \delta_{ij}$, $\langle H_{ij} H_{kl} \rangle_0 \sim
\delta_{il} \delta_{jk}$ so all equations like (\ref{selfenergy}) and (\ref{SigmaR}) reduce to scalar equations for the coefficients 
multiplying the delta functions.

\begin{figure}
\begin{center}
\includegraphics[width=16cm]{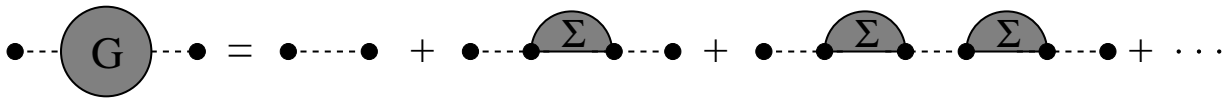}
\end{center}
\caption{Diagrams $G(z)$ can be obtained from one-line-irreducible
diagrams $\Sigma(z)$ (see figure \ref{Sexample}) by joining them one after another.
\label{GfromSigma}}
\end{figure}

\begin{figure}
\begin{center}
\includegraphics[width=5cm]{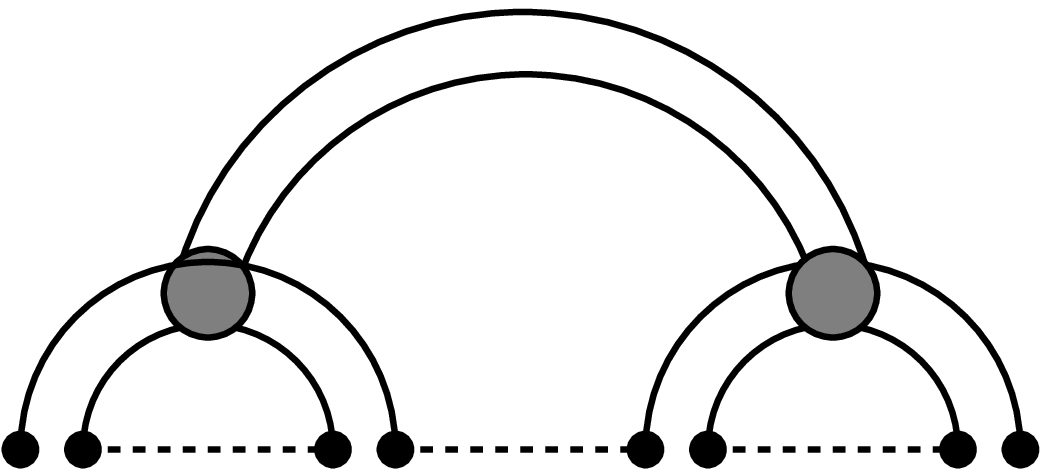} \qquad \qquad
\includegraphics[width=3cm]{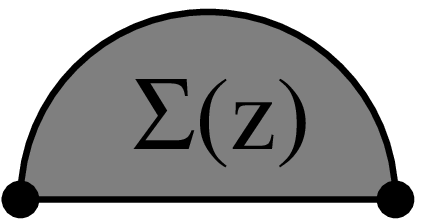}
\end{center}
\caption{(Right) An example of an one-line-irreducible diagram. 
(Left) The graphical notation for the generating function $\Sigma(z)$ of 
one-line-irreducible diagrams. 
\label{Sexample}}
\end{figure}

A graphical interpretation of equation (\ref{selfenergy})
becomes clear if one rewrites it as an infinite geometric series
\be
G(z) = \frac{1}{z} + \frac{1}{z} \Sigma(z) \frac{1}{z} + 
\frac{1}{z} \Sigma(z) \frac{1}{z} \Sigma(z) \frac{1}{z} + \ldots  \ .
\label{GSgs}
\ee
which can be seen in figure \ref{GfromSigma}.  
This figure tells us that all diagrams in $G(z)$
can be constructed by lining up one-line-irreducible diagrams one after another.
An example of such a one-line-irreducible diagram contributing to
$\Sigma(z)$ is shown in figure \ref{Sexample}. 
Such diagrams  are characterized by the property that they cannot be disconnected by cutting one line as opposed to diagrams generated by $G(z)$. Indeed, as one can see in
figure \ref{GfromSigma} a diagram in $G(z)$ can be disconnected by cutting 
any horizontal line like that between two consecutive $\Sigma$'s. The diagrammatic 
equation in figure \ref{GfromSigma} can be interpreted as a definition of the generating function $\Sigma(z)$ of one-line-irreducible diagrams. 

\begin{figure}
\begin{center}
\includegraphics[width=14cm]{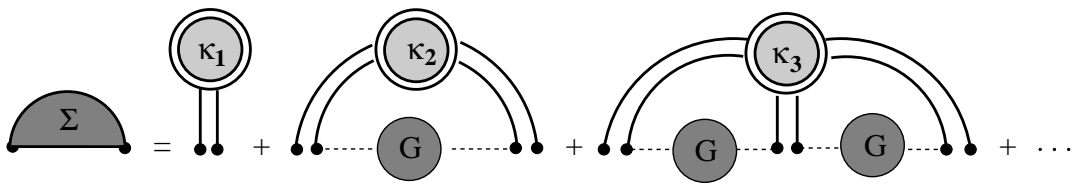}
\end{center}
\caption{A one-line-irreducible diagram can be obtained from a
one-line-reducible diagram by adding to it a minimal diagrammatic
structure complying with the measure (\ref{split}) which makes it
one-line-irreducible. Such a minimal structure is provided by diagrams corresponding 
to planar connected moments $\kappa_k$ (free cumulants) (see figure \ref{kth-conn}) which we indicated by bubbles surrounded by double circles in the figure. This double ring around the bubble is chosen to make it similar
to double brackets used in our notation for connected averages.
Diagrams in such a bubble are connected. The difference between 
diagrams corresponding to planar connected moments (cumulants)
and spectral moments is explained in figure \ref{kth-conn}. \label{SigmafromG}}
\end{figure}

It turns out that one can write down an independent equation relating
$\Sigma(z)$ to $G(z)$. One can namely observe that any one-line-irreducible diagram can be obtained from diagrams
generated by $G(z)$ as shown in figure \ref{SigmafromG} by adding a 
spider structure making them one-line-irreducible. 
Each bubble $\kappa_n$ of the spider with $n$ double legs 
corresponds to a connected moment (free cumulant) of order $n$.
This equation tells us that
\be
\Sigma(z) = \frac{1}{N} \langle\langle {\rm Tr} H \rangle\rangle + 
\frac{1}{N} \langle\langle {\rm Tr} H^2 \rangle\rangle G(z) + 
\frac{1}{N} \langle\langle {\rm Tr} H^3 \rangle\rangle G^2(z) + \ldots = R(G(z)) 
\ee
The diagrammatic equations in figures \ref{GfromSigma} and 
\ref{SigmafromG} belong to the category of Dyson-Schwinger 
equations known from quantum field theory. They are equivalent 
to the equations (\ref{selfenergy}) and (\ref{SigmaR}) 
discussed in the previous section. 
\begin{figure}
\begin{center}
\includegraphics[width=5cm]{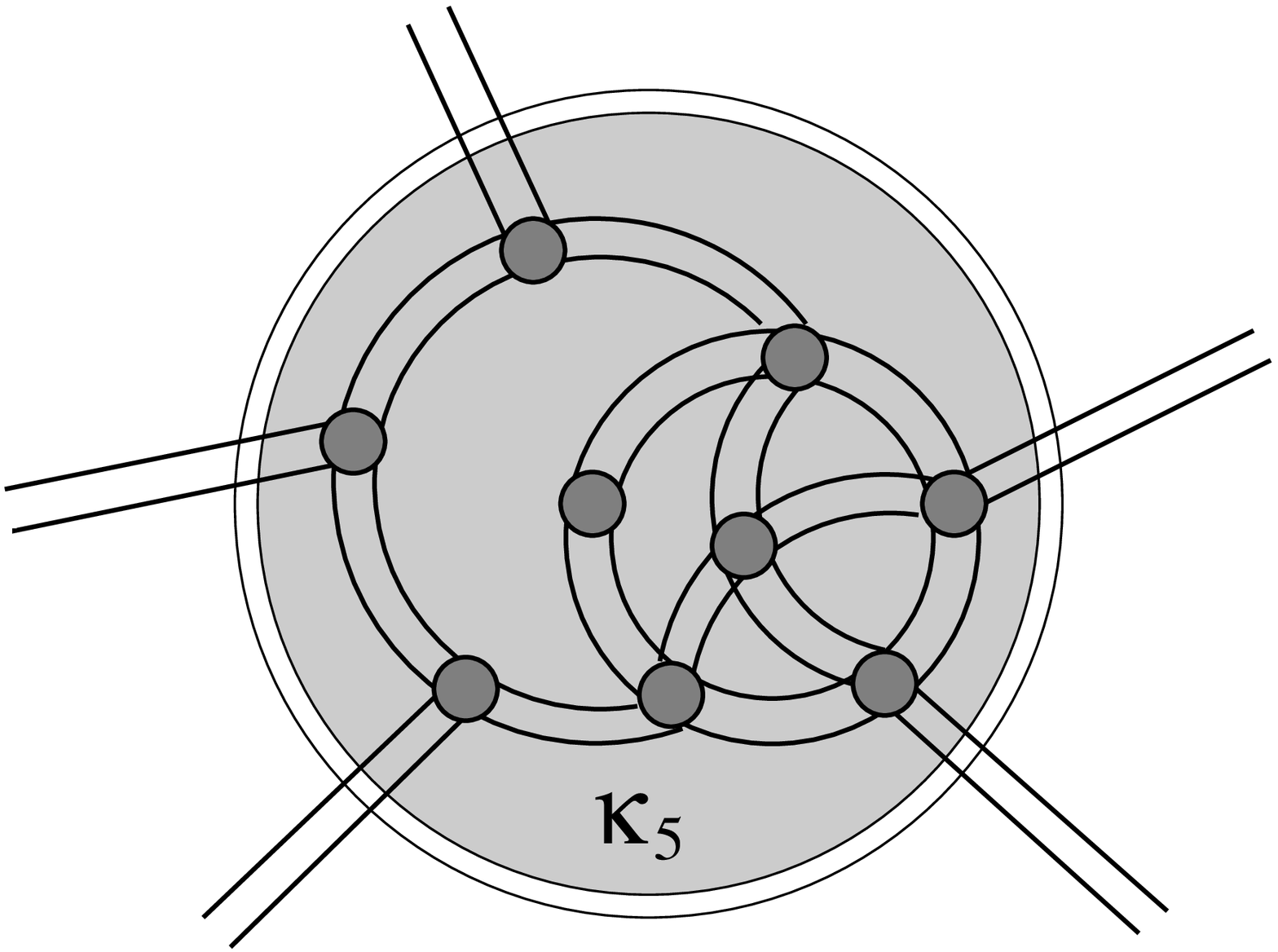} \qquad
\includegraphics[width=5cm]{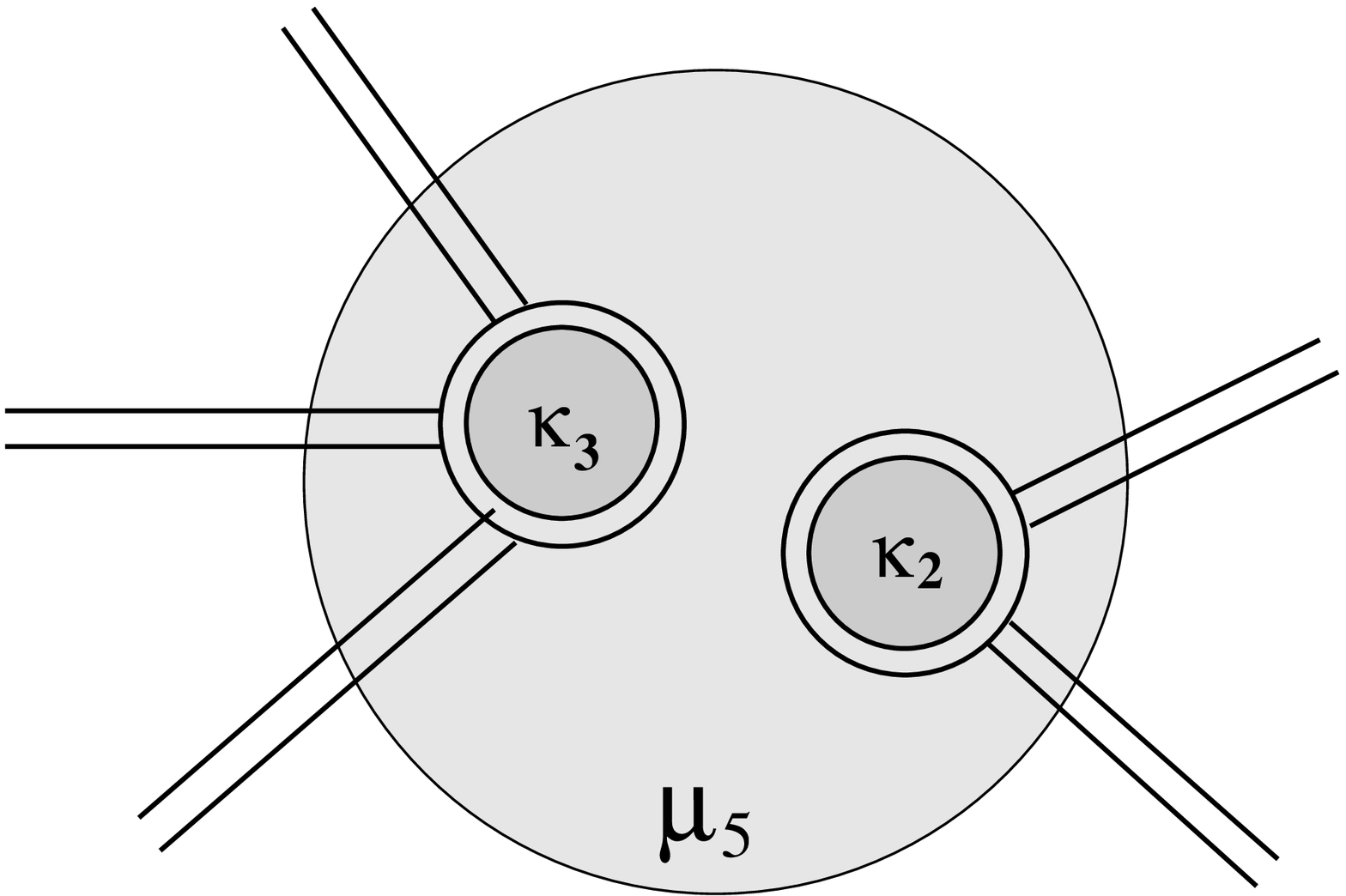} 
\end{center}
\caption{(Left) An example of a diagram generated by fifth free
cumulant $\kappa_5 = \frac{1}{N} \langle \langle {\rm Tr} H^5 \rangle\rangle$.
All diagrams in the bubble must be connected in contrast to the diagrams generated by spectral moments. (Right) An example of the decomposition of some diagrams generated by the fifth spectral moment $\mu_5 = \frac{1}{N} \langle {\rm Tr} H^5 \rangle$ into two connected moments $\kappa_2 \kappa_3$.
Some other diagrams in $\mu_5$ can be decomposed into 
$\kappa_1\kappa_2\kappa_2$ or any other combination of cumulants 
as long as the number of external legs is five. \label{kth-conn}
Only a small subset of diagrams in $\mu_5$ corresponds to those of $\kappa_5$.} 
\end{figure}

\subsection{Addition law: R transform}

The R transform~\cite{VOICULESCU} is important because it allows one to concisely
write down a law of addition of (free) independent large matrices.
Consider first a factorized measure for two large matrices $A,B$ 
in the limit $N\rightarrow \infty$
\be
P(A,B) = P_A(A) P_B(B)
\label{product_measure}
\ee
where $P_A(A) \sim \exp -N {\rm Tr} V_A(A)$ and $P_B(B) \sim \exp -N {\rm
Tr} V_B(B)$. 
Then consider a matrix $H=A+B$. The law of addition tells us how to
calculate the spectral density of $H$ for given spectral 
densities of $A$ and $B$. 

The idea is based on the observation that
connected planar moments (free cumulants) of the sum $H=A+B$ split
into two independent parts
\be
\frac{1}{N} \langle\langle {\rm Tr} (A+B)^k \rangle\rangle = 
\frac{1}{N} \langle\langle {\rm Tr} A^k \rangle\rangle + 
\frac{1}{N} \langle\langle {\rm Tr} B^k \rangle\rangle \ .
\ee
The reason for this separation of connected moments can be easily understood
in terms of Feynman diagrams. All mixed connected moments $\langle \langle {\rm Tr} A^{a} B^{b} A^c B^d ... \rangle \rangle$ disappear just because there is no direct line in a connected diagram between a vertex of type $A$ and $B$ since 
the $AB$-propagator is zero $\langle A_{ij} B_{kl} \rangle_0 = 0$. The crossed pairs of double lines corresponding to $A$ and $B$ vanish in the large $N$ limit, since they represent non-planar contribution. So all external lines of a bubble generated by $k$-th cumulant  correspond either
$A$ or to $B$.  In other words free cumulants fulfill a simple equation
\be
\kappa_{A+B,n} = \kappa_{A,n} + \kappa_{B,n}
\ee
and thus
\be 
R_{A+B}(z)=R_A(z)+R_B(z) \ .
\label{addition_law}
\ee
The argument given above is equivalent to a reasoning
based on non-crossing partitions used to prove this law in~\cite{SPEICHER}.
The law of free addition (\ref{addition_law}) 
is also sufficient to calculate spectral moments of the free sum
$\mu_n = 
\frac{1}{N} \langle {\rm Tr} H^n \rangle = \frac{1}{N}
\langle {\rm Tr} (A+B)^n \rangle$ if one knows 
the spectral moments of $A$ and $B$. 
The recipe follows from the 
relations (\ref{GfromR}) and (\ref{Rtransform}):
\begin{enumerate}
\item Using (\ref{Rtransform}) calculate $R_A(z)$ for given $G_A(z)$ 
and  $R_B(z)$ for $G_B(z)$. 
\item Construct the R transform $R_{A+B}(z)$ 
for the sum using the addition law (\ref{addition_law}). 
\item Calculate $G_{A+B}(z)$ for $R_{A+B}(z)$ using (\ref{Rtransform}) 
and calculate spectral moments $\langle{\rm Tr}(A+B)^n\rangle$ 
and the spectral density of $A+B$ using (\ref{recon}).
\end{enumerate}

\section{Non-hermitian Random Matrices \label{Sec_non_hermit}}

\subsection{Preliminaries}

We now briefly recall how to calculate the spectral density
of non-hermitian random matrices using generalized Green's functions~\cite{JANNOWPAPZAH}. 
The crucial difference between the hermitian and non-hermitian case 
comes from the fact that in non-hermitian random matrix models 
eigenvalues do not lie on the real axis. In the large $N$ limit
they may accumulate in {\em two-dimensional} domains in the complex plane 
and the corresponding eigenvalue density 
\be
\rho(z,\bar{z}) = 
\lim_{N\rightarrow \infty}
\frac{1}{N} \left\langle \sum_i \delta^{(2)}(z\!-\!\lambda_i)
        \right\rangle
\label{spect2-h}
\ee
may become a continuous function with an extended support 
in the complex plane. In particular, in stark contrast to the hermitian case,
the moments $\mu_n=\frac{1}{N} \langle {\rm Tr} X^n \rangle$ no longer
determine the eigenvalue density.
If one wants to apply the Green's function formalism
for (\ref{spect2-h}) one has to find a representation of the two-dimensional 
delta function and not as in the previous section of one dimensional one (\ref{spect-h}). A natural candidate is 
\be 
\delta^{(2)}(z-\lm_i)=\frac{1}{\pi} \lim_{\epsilon \rightarrow 0}
\frac{\epsilon^2}{(\epsilon^2 +|z-\lm_i|^2)^2} \ . 
\ee 
With help of this representation one can write
\be
\rho(z,\bar{z}) = 
\lim_{\epsilon \rightarrow 0} \lim_{N\rightarrow \infty}
\left\langle  \frac{1}{N} \sum_{i=1}^N 
\frac{\epsilon^2}{(\epsilon^2 +|z-\lm_i|^2)^2} \right\rangle
\label{poisson}
\ee
or 
\be
\rho(z,\bar{z}) = \frac{1}{\pi} \frac{\partial^2 F(z,\bar{z})}{\partial z\partial \bar{z}}
\ee
where
\be
F(z,\bar{z}) = \lim_{\epsilon\rightarrow 0} \lim_{N\rightarrow \infty}
\left\langle \frac{1}{N} \sum_{i=1}^N \ln 
\left( |z-\lambda_i|^2 + \epsilon^2 \right) \right\rangle 
\ee
or equivalently
\be
F(z,\bar{z}) = 
\lim_{\epsilon\rightarrow 0} \lim_{N\rightarrow \infty}
\left\langle \frac{1}{N} {\rm Tr } \ln 
\left( (z\mathbbm{1}-X)(\bar{z} \mathbbm{1}-X^{\dagger}) +
        \epsilon^2 \mathbbm{1} \right) \right\rangle \,.
\label{els} 
\ee 
One can interpret (\ref{poisson}) as a Poisson equation 
for electrostatics where $\rho(z,\bar{z})$ is a two-dimensional
charge distribution and $F(z,\bar{z})$ is a electrostatic potential
\cite{HAAKE,GIR,SOMMERS}. One can further exploit 
the electrostatic analogy by introducing the corresponding
electric field which is equal to
the Green's function 
 \be G(z, \bar{z})\!
\equiv\!
 \frac{\partial F}{\partial z}=
\lim_{\epsilon \rightarrow 0} \lim_{N\rightarrow \infty} 
\left\langle \frac{1}{N} {\rm Tr} \frac{\bar{z} \mathbbm{1}-X^{\dagger}}
{(\bar{z}\mathbbm{1}- X^{\dagger})(z \mathbbm{1} - X) +\epsilon^2 \mathbbm{1})}\right\rangle
. \label{GG} \ee 
up to a coefficient. $F$ is a real function on the complex plane, so it is a scalar field from the point of view of two-dimensional electrodynamics 
while $G$ is a complex function and a vector field, respectively.
The Poisson equation can be rewritten as a Gauss law in two-dimensions
\eq 
\label{e.dzbar} 
\rho(z,\zb) = \f{1}{\pi}\partial_{\zb} G(z,\zb)\ . 
\eqx
In the large $N$ limit when the eigenvalues $\lambda_i$ of the random matrix
coalesce in a certain region of the complex plane, the Green's function $G(z,\bar{z})$ is no longer holomorphic. Actually as one can see from the Gauss 
law (\ref{e.dzbar}) the eigenvalue distribution $\rho(z,\bar{z})$ is related to 
the non-holomorphic behavior of the Green's function. 

Let us make a few general remarks about the way we shall use
this electrostatic interpretation. In electrostatics one usually
applies the Gauss law to determine the electric field for a given
charge density. In our problem we proceed in the opposite direction.
We first calculate the Green's function (electric field) and 
then we use it to determine the eigenvalue density. 
Secondly, in order to calculate the average (\ref{GG}) one has to
take a double limit. It is important to take it in the correct 
order: first to send $N$ to infinity and only then $\epsilon$ to zero,
since if one took this limit in the opposite order by first setting
$\epsilon=0$ for a finite matrix, then the expression in the brackets in (\ref{GG}) would reduce to $1/N {\rm Tr} (z\mathbbm{1}-X)^{-1}$. Finally, whenever we apply generating functions for planar diagrams 
we can automatically take the limit $\epsilon\rightarrow 0$, which trivially amounts to setting $\epsilon=0$, since the large $N$ limit 
($N\rightarrow \infty$) has already been taken by the planar approximation
used to write relations between generating functions for planar diagrams.

Note that the Green's function (\ref{GG}) is a complicated object 
which does not resemble its hermitian counterpart -- in particular 
we cannot just apply the geometric series expansion that was 
crucial for calculations in the hermitian case (\ref{e.series}).
We can however use a trick, invented in \cite{JANNOWPAPZAH},
which allows us to apply the geometric series expansion but for an 
extended $2N \times 2N$ matrix: 
\be {{\cal {G}}}(z,\bar{z}) = 
\arr{{\GG}_{11}}{{\GG}_{1\bar{1}}}{{\GG}_{\bar{1}1}}
{{\GG}_{\bar{1}\bar{1}}} = 
\left\langle \frac{1}{N} 
\rm{Tr_{b2}} \setlength\arraycolsep{0pt} \arr{z\mathbbm{1}\!-\!X}{i
\epsilon\mathbbm{1}}{i \epsilon\mathbbm{1}}{\bar{z} \mathbbm{1}\!
-\!X^{\dagger}}^{-1}\right\rangle
\label{19}
\ee
where we have introduced the block-trace operation
\be
{\rm Tr_{b2}} \setlength\arraycolsep{3pt} \arr{X}{Y}{Z}{V}_{2N
\times 2N} \hspace*{-3mm}\equiv \setlength\arraycolsep{3pt} \arr{
{\rm Tr}\ X}{{\rm Tr}\ Y}{{\rm Tr}\ Z}{{\rm Tr}\ V}_{2 \times 2}
\hspace*{-5mm}\,. \label{block2} 
\ee  
which reduces $2N \times 2N$ matrices to $2\times 2$ ones. 
The elements of $\GG$ read explicitly:
\be
\GG_{11}(z,\bar{z}) &=& \left\langle \frac{1}{N} {\rm Tr} \frac{\zb \mathbbm{1} -X^{\dagger}}{(
          \zb \mathbbm{1} -X^\dagger)(z\mathbbm{1}-X)+\epsilon^2\mathbbm{1}} \right\rangle \nonumber \\
\GG_{1\bar{1}}(z,\bar{z}) &=& \left\langle \frac{1}{N} {\rm Tr} \frac{-i\epsilon \mathbbm{1} }{(z\mathbbm{1} -X)(\zb\mathbbm{1}-X^{\dagger})+\epsilon^2\mathbbm{1}} \right\rangle \nonumber \\
\GG_{\bar{1}1}(z,\bar{z}) &=& \left\langle \frac{1}{N} {\rm Tr} \frac{ -i\epsilon\mathbbm{1}}{(
          \zb\mathbbm{1} -X^\dagger)(z\mathbbm{1}-X)+\epsilon^2\mathbbm{1}} \right\rangle \nonumber \\
\GG_{\bar{1}\bar{1}}(z,\bar{z}) &=& \left\langle \frac{1}{N} {\rm Tr} \frac{z\mathbbm{1}-X}{(
          z\mathbbm{1} -X)(\zb\mathbbm{1}-X^{\dagger})+\epsilon^2\mathbbm{1}} \right\rangle
          \label{gggg}
\ee 
In all these
equations we tacitly assume the averages in the right hand side to be 
calculated in the double limit: first $N\rightarrow \infty$ 
and then $\epsilon \rightarrow 0$. The indices $11$, $1\bar{1}$
$\bar{1}1$ and $\bar{1}\bar{1}$ merely  reflect positions 
of blocks in the $2 \times 2$ matrix $\GG$. We see that the  
upper-right  $\GG_{11}$ is equal to the Green's function
$G(z,\bar{z}) = \GG_{11}(z,\bar{z})$ (\ref{GG}). On the other
hand, the main advantage of using the matrix $\GG$ is that it can be 
calculated using simple geometric series
expansion. Indeed, defining $2N \times 2N$ matrices 
\be 
\label{defzg}
{\ZZ_\epsilon}= \arr{z\mathbbm{1}}{i \epsilon \mathbbm{1}}{i
 \epsilon \mathbbm{1}}{\zb \mathbbm{1}}
\ee
and
\be
\label{defhg}
\HH=\arr{X}{0}{0}{X^{\dagger}} \,. 
\ee 
we can see that the generalized Green's
function is given formally by the same definition as the usual
Green's function $G$ but in the space of doubled dimensions 
\be 
\GG(z,\bar{z})= \lim_{\epsilon \rightarrow 0} 
\lim_{N\rightarrow \infty} \frac{1}{N} \left\langle {\rm Tr_{b2}}
\frac{1}{{\ZZ_\epsilon }-\HH}\right\rangle
        \,.
\label{conciseZ} 
\ee 
For the sake of the argument we have written now the double limit explicitly. 
As in the hermitian case, the Green's function
is completely determined by the knowledge of `generalized' moments.
They are now however matrix-valued
\be \lim_{\epsilon \rightarrow 0} 
\lim_{N\rightarrow \infty} \frac{1}{N}
\left\langle {\rm Tr_{b2}}\,\, \ZZ_\epsilon^{-1} \HH \ZZ_\epsilon^{-1}
 \HH \ldots \ZZ_\epsilon^{-1}
        \right\rangle
\label{genmom} 
\ee 
and are not easily related to the eigenvalue density.
As before, we now proceed by applying the diagrammatic techniques to 
determine the non-hermitian Green's function. 
We begin by writing equations for 
generating functions for planar diagrams.
In analogy to (\ref{selfenergy}), we introduce the self-energy $\Sigma$ 
but now as a matrix-valued function
\be {\Sigma}(z,\bar{z})
&\equiv& \setlength\arraycolsep{3pt}
\arr{\Sigma_{11}(z,\zb)}{\Sigma_{1\bar{1}}(z,\zb)}
{\Sigma_{\bar{1}1}(z,\zb)}{\Sigma_{\bar{1}{\bar{1}}}(z,\zb)}
\ee
As in the hermitian case $\Sigma$ is a generating function  
for one-line irreducible diagrams. In general it is not diagonal.
Formally it is related to the Green's function as
\be\GG(z,\bar{z})=
\left({\ZZ}-{\Sigma}(z,\bar{z})\right)^{-1} . \label{gin1} 
\ee 
where $\ZZ$ is a diagonal $2 \times 2$ matrix
\be {\cal Z}= \arr{z}{ 0 }{0}{\zb}
\label{diagZ}
\ee 
obtained from $\ZZ_\epsilon$ by taking 
block trace and setting $\epsilon=0$. This may be done
since the equation (\ref{gin1}) is already in the limit 
$N\rightarrow \infty$. From here on we will set $\epsilon=0$
in all equations.

An explicit solution for the Green's function $G(z,\bar{z}) = \GG_{11}(z,\bar{z})$ takes
therefore the following 
form 
\be 
G(z,\bar{z}) =\frac{\zb-\Sigma_{\bar{1}\bar{1}}}{(z-\Sigma_{11})(\zb-\Sigma_{\bar{1}\bar{1}})-\Sigma_{1\bar{1}}\Sigma_{\bar{1}1}}
\label{importantdet} \,. 
\ee
We skipped arguments $(z,\bar{z})$ of $\Sigma$'s on the right hand side to
shorten the notation. 
The non-diagonal terms in (\ref{19}) also contain an interesting 
information~\cite{NOERENBERGPAPER}, namely their product is equal to the correlator 
between left $(\langle L_i|) $ and right $(|R_i\rangle)$  eigenvectors of $X$, introduced 
originally in~\cite{CHALKERMEHLIG}
\be
C(z,\zb) \equiv -\GG_{1\bar{1}}\GG_{\bar{1}1}=\frac{\pi}{N}\left\langle    \sum_{i=1}^N \langle L_i|L_i\rangle\langle R_i|R_i\rangle\delta^{(2)}(z-\lambda_i)  \right\rangle
\label{Cdef}
\ee 
Since $C(z,\zb)$ vanishes outside the eigenvalue support, and for typical nonhermitian
ensembles is nonzero, the condition $C(z,\zb)=0$ often provides 
a convenient equation for the boundary separating holomorphic and nonholomorphic solutions of the spectral problem. Indeed,  when off-diagonal terms of $\Sigma$ vanish 
 equation (\ref{importantdet}) simplifies to that for hermitian matrices
$G = 1/(z-\Sigma_{11})$.

As in the hermitian case we can write an independent equation 
relating $\GG$ and $\Sigma$ -- a counterpart of (\ref{SigmaR}).
The R transform however is now a more complicated object since
it maps a $2 \times 2$ matrix $\cal G$ onto
a $2 \times 2$ matrix $\Sigma$:
\be
\Sigma(z,\bar{z}) = {\cal R}\left({ \cal G}(z,\bar{z})\right)
\label{SR_nh}
\ee
or in an explicit notation 
\be
\left(
\begin{array}{cc} 
\Sigma_{11}(z,\bar{z}) & \Sigma_{1\bar{1}}(z,\bar{z}) \\
\Sigma_{\bar{1}1}(z,\bar{z}) & \Sigma_{\bar{1}\bar{1}}(z,\bar{z})
\end{array}
\right) = 
\left(
\begin{array}{cc} 
{\cal R}_{11}\left(\GG(z,\bar{z})\right) & 
{\cal R}_{1\bar{1}}\left(\GG(z,\bar{z})\right) \\
{\cal R}_{\bar{1}1}\left(\GG(z,\bar{z})\right) & 
{\cal R}_{\bar{1}\bar{1}}\left(\GG(z,\bar{z})\right) 
\end{array}
\right)
\ee
In order to complete the analogy to the hermitian case we shall now
provide a diagrammatic interpretation of the last relation.

\subsection{Planar Feynman diagrams for non-hermitian matrices}

\begin{figure}
\begin{center}
\includegraphics[width=11cm]{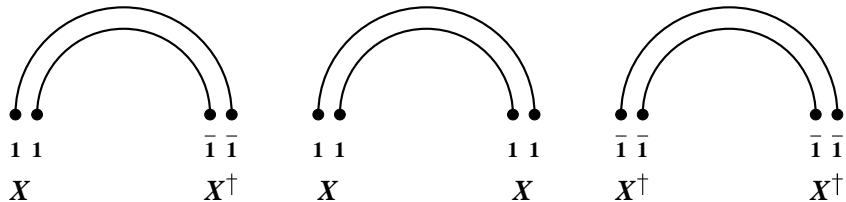} \qquad
\caption{Propagators for non-hermitian models generated
by the Gaussian part of the measure (\ref{P0}). The are 
obtained by identification $X \leftrightarrow \HH_{11}$ and $X^{\dagger} \leftrightarrow \HH_{\bar{1}\bar{1}}$ (\ref{defhg}). This identification
induces the indexing of the endpoints marked as dots in the figure
(Left) 
$\left\langle XX^\dagger\right\rangle_0=\left\langle \HH_{11} \HH_{\bar{1}\bar{1}}\right\rangle_0= \sigma^2$;
(Middle) 
$\left\langle X X \right\rangle_0 = \left\langle \HH_{11} \HH_{11}\right\rangle_0 = \tau \sigma^2$;
(Right) $\left\langle X^\dagger X^\dagger \right\rangle_0=\left\langle \HH_{\bar{1}\bar{1}} 
\HH_{\bar{1}\bar{1}}\right\rangle_0 = \tau \sigma^2$.
\label{dprop}}
\end{center}
\end{figure}

\begin{figure}
\begin{center}
\includegraphics[width=9cm]{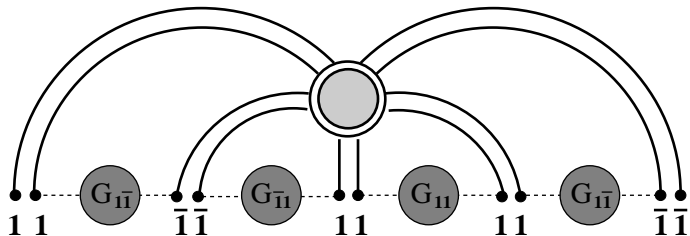} 
\caption{Connected diagrams  generated by
the fifth order planar cumulant 
$\langle\langle XX^\dagger XXX^\dagger\rangle\rangle$ (the head of the spider). 
These diagrams contribute 
a factor $\GG_{1\bar{1}} \GG_{\bar{1}1} \GG_{11} \GG_{1\bar{1}}$ to
$\Sigma_{1\bar{1}} = \RR_{1\bar{1}}(\GG)$.
\label{spider5}}
\end{center}
\end{figure}

We shall now discuss the diagrammatic method of calculating 
eigenvalue densities for non-hermitian random matrices generated by probability 
measures of the type $P(X) \sim \exp\left( -N {\rm Tr} V(X,X^\dagger)\right)$
in the limit $N\rightarrow \infty$, which as before corresponds to
the limit of planar diagrams. We  consider potentials given by sums
of terms being alternating sequences of powers of $X$ and $X^ \dagger$ 
like $X X^\dagger X^2 X^\dagger \ldots $. Such a potential must be
hermitian $\left[V(X,X^\dagger)\right]^\dagger = V(X,X^\dagger)$ to ensure that 
the expression in the exponent is a real number. 
The first step of the diagrammatic construction it to split the measure 
into the Gaussian part and the residual one $P(X) = P_0(X) P_r(X)$ and use $P_0(X)$ to calculate averages $\langle \ldots \rangle_0$ which can be represented
as Feynman diagrams, exactly as for hermitian matrices (\ref{split}). 
The Gaussian measure $P_0(V) \sim e^{-N {\rm Tr} V_0(X)}$
is constructed from a quadratic potential. The most general form
of a quadratic potential being a real number 
is ${\rm Tr} V_0(X) = a {\rm Tr} XX^\dagger + b {\rm Tr}\left(X^2 + X^{\dagger 2}\right)$ with some real coefficients $a,b$. 
The coefficients must be appropriately chosen to ensure the potential
be positive. The expresion is manifestly positive when expressed 
in new parameters $\sigma,\tau \in (-1,1)$:
\be
P_0(X) \sim \exp \left\{ - N \frac{1}{\sigma^2} \frac{1}{1-\tau^2} \left( 
{\rm Tr} XX^\dagger - \tau \frac{1}{2} {\rm Tr} \left(XX +X^\dagger X^\dagger\right)\right)\right\} 
\label{P0}
\ee
as one can see for example by writing down the corresponding
two-point correlation functions (propagators):
\be
\nonumber
\left\langle X_{ab} X^\dagger_{cd} \right\rangle_0 =
\left\langle X^\dagger_{ab} X_{cd} \right\rangle_0 = \frac{\sigma^2}{N} \delta_{ad} \delta_{bc} \ , \\
\big\langle X_{ab} X_{cd} \big\rangle_0 = \left\langle X^\dagger_{ab} X^\dagger_{cd} \right\rangle_0 =
\tau \cdot \frac{\sigma^2}{N} \delta_{ad} \delta_{bc} 
\label{tau_prop}
\ee
The propagators represent elementary building blocks of Feynman
diagrams. As for hermitian matrices the propagators are
proportional to delta functions, so after taking the block trace
we can reduce the problem to a $2 \times 2$ one with propagators
corresponding to $XX^\dagger$, $X^{\dagger}X$, $XX$, $X^\dagger X^\dagger$.
The crucial step in inferring the index structure of 
equations relating $2 \times 2$ matrices $\GG$ and $\Sigma$ is to use
the correspondence between $X \leftrightarrow \HH_{11}$ and 
$X^\dagger \leftrightarrow \HH_{\bar{1}\bar{1}}$ 
which follows from equation (\ref{defhg}). Let us do that.
The two-point functions (\ref{tau_prop}) reduce to
propagators represented by double arcs shown in figure \ref{dprop}.
The matrix ${\cal Z}^{-1}$ (\ref{diagZ}) generates lines 
between $11$ vertices which contribute $1/z$ and
lines between $\bar{1}\bar{1}$ which contribute $1/\bar{z}$, while
there are no lines between mixed vertices. Using these elementary blocks 
we can draw graphical equations as those in figures \ref{GfromSigma} and \ref{SigmafromG}. The only difference as compared to
the hermitian case is that they are written for  $2 \times 2$ 
matrices. Each black dot in the diagrams in these figures 
is ascribed to an index which may assume two values: either $1$ or $\bar{1}$. 
Each pair of neighboring dots on the horizontal line in figure \ref{SigmafromG} corresponds to $X$ or $X^\dagger$ or to 
$\HH_{11}$ or $\HH_{\bar{1}\bar{1}}$ as follows from the assignment (\ref{defhg}). As an example consider a spider diagram of order five 
generated in the expansion shown in figure \ref{SigmafromG}. Each leg of the spider may be attached to $X$ or $X^\dagger$, so on the horizontal line we have a sequence of these symbols -- for instance $XX^\dagger XXX^\dagger$, or equivalently $\HH_{11} \HH_{\bar{1}\bar{1}} \HH_{11} \HH_{11} \HH_{\bar{1}\bar{1}}$ (\ref{defhg}). The corresponding diagram is shown in figure \ref{spider5}.
In a shorthand notation the diagram is determined by a sequence of pairs
$11,\bar{1}\bar{1},11,11,\bar{1}\bar{1}$ on the horizontal line which 
begins with $1$ and ends with $\bar{1}$ so it contributes to $\Sigma_{1\bar{1}}$, since the corresponding diagram is one-line irreducible. 
As one can see from the figure its contribution is proportional to
$\GG_{1\bar{1}} \GG_{\bar{1}1} \GG_{11} \GG_{1\bar{1}}$. 
The indices of $\GG$ bubbles are enforced by indices of the spider legs -- 
they must match the sequence on the horizontal line. 

All such contributions are captured by a matrix valued 
function $\RR(\GG)$, in this particular
case by its element $\RR_{1\bar{1}}(\GG)$ which contains contributions
generated by sequences beginning with $1$ and ending with $\bar{1}$.
Each element of the matrix $\RR(\GG)$ may depend on all elements
of the matrix $\GG$ so this function maps $2\times 2$ matrices onto 
$2\times 2$ matrices and in general is highly nontrivial~(\ref{SR_nh}). 
The exception is the Gaussian case for which the map is linear. 

For the purpose of this paper let us study Gaussian case 
in more detail. The most general Gaussian ensemble (\ref{P0}) leads through 
(\ref{tau_prop}) to (see figure~\ref{dprop})
 \be
 \RR(\GG)=\left(
 \begin{array}{cc}
  \Sg_{11} & \Sg_{1\bar{1}}\\ \Sg_{\bar{1}1}& \Sg_{\bar{1}\bar{1}} \end{array} \right) =
  \left(
 \begin{array}{cc}
  \tau \sg^2 \GG_{11} & \sg^2 \GG_{1\bar{1}} \\ 
  \sg^2 \GG_{\bar{1}1} & \tau \sg^2 \GG_{\bar{1}\bar{1}} \end{array} \right) 
  \label{general}
  \ee
Let us now constrain ourselves to the so-called Ginibre-Girko ensemble which 
corresponds to the case $\tau=0$ and $\sg=1$ in (\ref{dprop}), 
so the matrix $\Sigma$ reads 
 \be
\RR(\GG)= \left(
 \begin{array}{cc}
  \Sg_{11} & \Sg_{1\bar{1}}\\ \Sg_{\bar{1}1}& \Sg_{\bar{1}\bar{1}} \end{array} \right) =
  \left(
 \begin{array}{cc}
  0 & \GG_{1\bar{1}} \\ \GG_{\bar{1}1} & 0 \end{array} \right) 
  \label{gin2}
  \ee
where the off-diagonal contributions are analogous to the relation $R(G)=G$ for the hermitian Gaussian ensemble.  
Solving~(\ref{gin1}-\ref{gin2})  determines the spectral problem for the Ginibre-Girko ensemble. Inserting (\ref{gin2}) into (\ref{gin1})
we get:
 \be
 \left(
 \begin{array}{cc}
  \GG_{11} & \GG_{1\bar{1}}\\ \GG_{\bar{1}1}& \GG_{\bar{1}\bar{1}} \end{array} \right) =
 \frac{1}{|z|^2-\GG_{1\bar{1}}\GG_{\bar{1}1}} \cdot  \left(
 \begin{array}{cc}
  \bar{z} & \GG_{1\bar{1}} \\ \GG_{\bar{1}1} & z \end{array} \right) 
  \label{gin3}
  \ee
The equation for off-diagonal element  reads
  \be
  \GG_{1\bar{1}}=\frac{\GG_{1\bar{1}}}{|z|^2-\GG_{1\bar{1}}\GG_{\bar{1}1}} \ .
  \ee
It has two-solutions: one with $\GG_{1\bar{1}}=0$ and the another one with $\GG_{1\bar{1}} \neq 0$. 
  The first one leads to a holomorphic Green's function $G=\GG_{11}$
  \be
  G(z)=\frac{1}{z}
  \ee
while the second one to a non-holomorphic (see the
upper diagonal component of equation $G\equiv \GG_{11}$ (\ref{gin3}))
  \be
  G(z,\bar{z})=\bar{z}
  \ee
  which gives  the following eigenvalue density 
  \be
  \rho(x,y)=\frac{1}{\pi} \frac{\partial}{\partial\zb}\GG_{11}(z,\zb)=\frac{1}{\pi} \ .
  \ee
  Both solutions match at the boundary $|z|^2=1$. 
  So we have recovered a known result that the  complex eigenvalues of the Ginibre-Girko ensemble are uniformly distributed on the unit disc.  
  
\subsection{Addition law for non-hermitian matrices}

One can actually use exactly the same arguments as for hermitian 
matrices to deduce the law of free addition for non-hermitian matrices.
It has a simple form given in terms of matrix-valued R transforms:
\be
{\cal R}_{A+B}(\GG) = {\cal R}_A(\GG) + {\cal R}_B(\GG)
\ee
which follows from the fact that all mixed $AB$ propagators vanish
and therefore all mixed connected diagrams having a line
between $A$ and $B$ vanish too. Since such diagrams represent
connected moments (free cumulants), e.g. $\frac{1}{N}\langle\langle AB^2A^\dagger AB^\dagger\rangle\rangle=0$, we see that the 
only non-zero contributions come from connected diagrams (moments)
which either have all $A$'s or all $B$'s. For applications and more details  of this generalized addition law we refer to~\cite{JANNOWPAPZAH,FEIZEE,CHAWAN}. 

\section{Multiplication law \label{Sec_multip}}

\subsection{Preliminaries \label{SRrelations}}

The S transform plays the same role for matrix multiplication
as the R transform for addition. Assume that $A$ and $B$ are 
large independent (free) random  matrices given by a 
product measure (\ref{product_measure}). 
The multiplication law tells us how to calculate
spectral moments $\frac{1}{N} \langle {\rm  Tr} (AB)^n\rangle$ of the 
product $H=AB$  provided we know the spectral moments of $A$ and $B$
or equivalently that we know the corresponding Green's functions
$G_A(z)$ and $G_B(z)$. The multiplication law, expressed in terms of
the S transform, reads~\cite{VOICULESCU}
\be 
S_{A\cdot B}(z)=S_A(z)S_B(z)
\label{law18} 
\ee 
and the S transform is defined by
\be S(z) = \frac{1+z}{z}\chi(z),  \,\,\,\, {\rm where}\,\,\,\,\,
\chi\left(zG(z)-1\right)=\frac{1}{z}  \ . \label{Stransform} 
\ee 
The algorithm for "multiplication" is similar to that for "addition":\\
(i) Calculate $S_A(z)$ and $S_B(z)$ using (\ref{Stransform}).\\
(ii) Use the multiplication law (\ref{law18}).\\
(iii) Use again~(\ref{Stransform}) to derive $G_{AB}(z)$
for the product of $AB$. 

Let us first derive  some useful relations between
the R and S transforms. Changing variables 
$z= y G(y) - 1$ in (\ref{Stransform}) we get
\be
S(y G(y) - 1) = \frac{1}{y - \frac{1}{G(y)}} \  .
\ee
Using (\ref{selfenergy}) we can rewrite the last equation as
\be
S\left(G(y) \Sigma(y)\right) = \frac{1}{\Sigma(y)} \ .
\ee
Setting $\Sigma(z) = R(G(z))$ and taking the reciprocals of  both sides we arrive at
\be
\frac{1}{S\left(G(y) R(G(y))\right)} = R(G(y)) \ .
\ee
Changing variables once again to $z=G(y)$ we  obtain the  equation
\be
R(z) = \frac{1}{S\left(z R(z)\right)}
\label{RS}
\ee
which gives an explicit relation between the R and S transforms.
The S transform can be defined only if the R trasform does 
not vanish at the origin: $R(0) \ne 0$. 
This corresponds to random matrices with a non-vanishing first moment
(cumulant) $\frac{1}{N} \langle {\rm Tr} H\rangle =
\frac{1}{N} \langle\langle {\rm Tr} H\rangle\rangle \ne 0$. 
Otherwise the S transform cannot be defined as a power series and all the manipulations
presented above break down. The last equation can be inverted. 
Let us introduce a new variable $y=z R(z)$. Now (\ref{RS}) reads
\be
S(y) = \frac{1}{R\left(\frac{y}{R(z)}\right)} = \frac{1}{R\left(\frac{y}{R\left(\frac{y}{R(z)}\right)}\right)} =  
\frac{1}{R\left(\frac{y}{R\left(\frac{y}{R(\ldots)}\right)}\right)}
\label{cont_frac}
\ee
where $z$ can be recursively eliminated by
repeating the substitution $z=\frac{y}{R(z)}$ ad infinitum. This
leads to a function which is nested infinitely many times forming a sort
of continued fraction. The last equation can alternatively be written as
\be
S(z) = \frac{1}{R\left(z S(z)\right)}
\label{SR}
\ee
which is an inverse formula to (\ref{RS}). The two equations 
can written in a symmetric way as mutually inverse maps
\be
z = y S(y) \quad \mbox{and} \quad y = z R(z) \ .
\ee
As an example we consider a shifted Gaussian random matrix which 
has only two first non-vanishing cumulants. For the standardized choice
$\kappa_1=\kappa_2=1$ the R transform reads $R(z) =1 + z$.
Using (\ref{SR}) we obtain
\be
S(z) = \frac{1}{1 + zS(z)}
\ee
and hence $S(z) = \frac{-1 + \sqrt{1+4z}}{2z}$.

\subsection{Diagrammatic derivation of the multiplication law}

We are now ready to diagrammatically derive 
the S transform and the corresponding multiplication law.
The argument given below will turn out to be  crucial for the generalization 
to non-hermitian matrices. The initial point of the construction is to consider a $2N \times 2N$ block matrix $\HH$ and its even powers\footnote{This should not be confused with the $2N\times 2N$ block matrix constructed for the nonhermitian Random Matrix Ensembles in section IV.}
\be
\HH=\arr{0}{A}{B}{0} \ , \quad \HH^{2k}=\arr{(AB)^k}{0}{0}{(BA)^k} \, .
\label{HH}
\ee
The upper-left corner of $\HH^{2k}$ involves solely the  powers of $AB$, 
which we are interested in. In order to have an access to the traces of individual blocks in the matrix we again apply the block trace operation defined before. 
The upper-left corner of the reduced matrix ${\rm Tr_{b2}} \HH^{2k}$ 
is equal ${\rm Tr} (AB)^k$ while of ${\rm Tr_{b2}} \HH^{2k+1}$ is equal
zero. So now the idea is to reformulate the problem of calculating
the Green's function for the product
\be
G_{AB}(z)=\frac{1}{N} \left\langle {\rm Tr} \frac{1}{z \mathbbm{1} -AB} \right\rangle
\ee
as a problem of calculating the upper-left corner of the Green's 
function $\GG(w)$ for the matrx $\HH$:
\be
\GG(w)= \left( 
\begin{array}{ll} \GG_{11}(w) & \GG_{12}(w) \\
                  \GG_{21}(w) & \GG_{22}(w)
\end{array} \right) = 
\frac{1}{N} \left\langle {\rm Tr_{b2}} \frac{1}{w \mathbbm{1}-\HH}\right\rangle \ .
\label{concise}
\ee
where $w$ is a complex number and $\mathbbm{1}$ is a unity matrix 
of dimensions $2N \times 2N$.
One can easily check that 
\be 
G_{AB}(z=w^2)=\frac{\GG_{11}(w)}{w}
\label{GabG11}
\ee 
since only every second (even) power of $\GG(w)$ 
contributes to the power expansion of $\GG_{11}(w)$,
which is thus a power expansion in $z=w^2$.

The next step is to define self-energy ${\Sigma}(w)$. It is 
a $2\times 2$ matrix
\be
\Sigma(w)
&\equiv& \setlength\arraycolsep{3pt}
\left(
\begin{array}{rr}
\Sigma_{11}(w) & \Sigma_{12}(w) \\
\Sigma_{21}(w) & \Sigma_{22}(w)
\end{array}
\right)
\label{Sigma22}
\ee
that is related to the Green's function as 
\be
\GG(w) = \left(w\mathbbm{1} - \Sigma(w)\right)^{-1} 
\label{DS1}
\ee
in analogy to (\ref{selfenergy}). All the matrices in the last equation
are of $2\times 2$ dimensions. This is the first Dyson-Schwinger 
equation. To write down the second one -- a counterpart of (\ref{SigmaR}) it is convenient to use its diagrammatic representation as that in figure \ref{SigmafromG}.
Instead of a scalar equation (\ref{SigmaR}) 
we will have a matrix equation for $2\times 2$ matrices 
$\GG$ and $\Sigma$ (\ref{concise}) and (\ref{Sigma22}).
Since we have now have $2\times 2$ matrices it is crucial
to work out the index structure of the corresponding equation.
This structure stems from the correspondence
$A \leftrightarrow \HH_{12}$ and $B \leftrightarrow \HH_{21}$ that 
follows from the position of the blocks in $\HH$ (\ref{HH}). 
Note the difference to the case discussed in the previous section were
we had diagonal blocks  (\ref{defhg}). 

\begin{figure}
\begin{center}
\includegraphics[width=14cm]{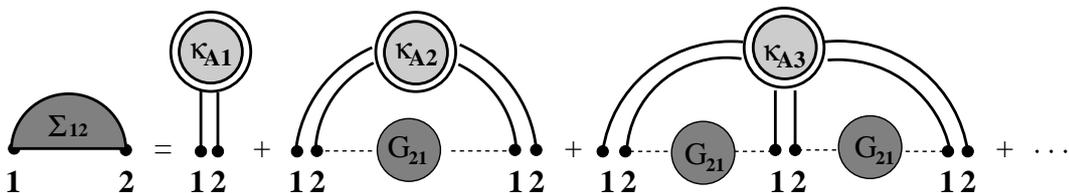}
\end{center}
\caption{
\label{RA_G21}
The spider diagrams correspond to free cumulants 
generated by the matrix $A = {\cal H}_{12}$ and therefore the double dots 
on the horizontal line are indexed by $12$. So on the horizontal line we have
alternating indices $1212\ldots 12$ and this enforces all the 
$\GG$-bubbles to have indices $21$ as one can see in the figure. 
Therefore there are only ${\cal G}_{21}$ bubbles in the diagram and
the corresponding equation is $\Sigma_{12}(w) = R_A({\cal G}_{21})$. 
The analogous equation for $B$-cumulants is $\Sigma_{21}(w) = R_B({\cal G}_{12})$. 
Similarly, one can see that $\Sigma_{11}(w)=\Sigma_{22}(w)=0$ since one of the
double dots on the horizontal line would need to have identical indices,
for which as we know from (\ref{HH}) the double line is equal zero.}
\end{figure}

The only non-vanishing cumulants are
$\frac{1}{N} \langle \langle {\rm Tr} {\cal H}_{12}^n \rangle\rangle=
\frac{1}{N}  \langle \langle {\rm Tr} A^n \rangle\rangle \equiv \kappa_{A,n}$ or 
$\frac{1}{N} \langle \langle {\rm Tr} {\cal H}_{21}^n \rangle\rangle= \frac{1}{N}  \langle \langle {\rm Tr} B^n \rangle\rangle \equiv \kappa_{B,n}$ while all mixed ones vanish as we discussed in the previous section. Due to this, the index structure of non-vanishing one-line-irreducible diagrams is restricted to that shown in figure \ref{RA_G21}
and its counterpart obtained by exchanging $1\leftrightarrow 2$ and
$A \leftrightarrow B$.
The diagrammatic equations discussed in figure \ref{RA_G21}
can be summarized as 
\be 
\Sigma(w) = \left( \begin{array}{cc} 
0 & R_A(\GG_{21}(w)) \\
R_B(\GG_{12}(w)) & 0 
\end{array}\right) \ .
\label{DS2}
\ee
Insering this into (\ref{DS1}) yields
\be
\left(
\begin{array}{cc} 
\GG_{11}(w)  & \GG_{12}(w) \\
\GG_{21}(w)  & \GG_{22}(w) 
\end{array} 
\right) =
\left(
\begin{array}{cc} 
w & - R_A\left(\GG_{21}(w)\right) \\
- R_B\left(\GG_{12}(w)\right)  & w  
\end{array} 
\right)^{-1} 
\label{DS}
\ee
which gives a direct relation between the Green's function
$\GG(w)$ and the R transform. 

We shall now rewrite this equation in a way which explicitly exhibits multiplicative
structure. Note that in the following manipulations we do not need to assume
anything about the first moment i.e. whether the ensemble is centered or not.
Inverting the matrix on the right hand side we obtain
\be 
\GG_{12}(w)= \frac{1}{\rm Det} R_A(\GG_{21}(w)) \ , \quad
\GG_{21}(w)= \frac{1}{\rm Det} R_B(\GG_{12}(w)) 
\label{GG_GG}
\ee
\be
\GG_{11}(w)=\GG_{22}(w)= \frac{w}{\rm Det} \ . 
\ee 
where ${\rm Det}$ is the determinant of the matrix, 
$w\mathbbm{1}-{\Sigma}(w)$, on the right hand side of (\ref{DS}):
\be
{\rm Det} = w^2 - R_A\left(\GG_{21}(w)\right) R_B\left(\GG_{12}(w)\right) \ .
\ee
Inserting two last equations to (\ref{GabG11}) we obtain
\be
G_{AB}(z)= \frac{\GG_{11}(w)}{w} = \frac{1}{\rm Det} = \frac{1}{z-R_A\left(\GG_{21}(w)\right) R_B\left(\GG_{12}(w)\right)}
\label{Gg}
\ee 
where $z=w^2$. Comparing the denominator in this equation to that
of the standard equation $G_{AB}(z) = 1/(z-R_{AB}(G_{AB}(z))$ (\ref{GfromR}) 
we get 
\be
R_{AB}\left(G_{AB}(z)\right) = 
R_A\left(\GG_{21}(w)\right) R_B\left(\GG_{12}(w)\right) \ .
\label{RRR}
\ee
At this stage we see the first hint of a multiplicative structure emergence. In order to complete this equation we also need (\ref{GG_GG}).
Let us set $g=G_{AB}(z)$, $g_A=\GG_{12}(w)$ and $g_B=\GG_{21}(w)$ to simplify
arguments in the R transforms in the last equation. Using this
substitution we can write (\ref{RRR}) and (\ref{GG_GG}) in a compact form 
as a closed set of equations for the R transform of the product
\be
R_{AB}(g) = R_A(g_B) R_B(g_A)
\label{main1}
\ee
and
\be
g_A = g R_A(g_B) \ , \quad g_B = g R_B(g_A) \ .
\label{main2}
\ee
which is equivalent to (\ref{R_RR}) announced at the beginning of the paper.
This is the multiplication law formulated in terms of the R transform.
Its main advantage in comparision to the S transform is that it can be applied 
even to centered ensembles (i.e. having vanishing mean) including the case
when both are centered (see the examples in sections VIA and VIB).

The difference with respect to the conventional multiplication law 
$S_{AB}(z)=S_A(z) S_B(z)$
is that the individual factors appearing in (\ref{main1}) are not expressed 
uniquely in terms of the properties of a \emph{single} random matrix ensemble 
e.g. the factor $R_A(\cdot)$ is evaluated on $g_B$ which is related to the ensemble $B$. However it is straightforward to obtain from 
(\ref{main1})-(\ref{main2}) the conventional multiplication law as we shall
illustrate below. 

Let us introduce a new variable $y = g R_{AB}(g)$. We can now
express $g_B$ -- the argument of $R_A$ purely in terms of the properties of
ensemble $A$:
\be
g_B = g R_B(g_A) = g \frac{R_{AB}(g)}{R_A(g_B)} = \frac{y}{R_A(g_B)} =
\frac{y}{R_A\left(\frac{y}{R_A(\ldots)}\right)} 
\ee
Now each of the factors in (\ref{main1}) depends on a single ensemble. We may do 
the same for the left hand side, which becomes (\ref{cont_frac}) 
\be
R_{AB}\left(\frac{y}{R_{AB}(g)}\right) = R_{AB}\left(\frac{y}{R_{AB}\left(\frac{y}{R_{AB}(\ldots)}\right)}\right) = \frac{1}{S_{AB}(y)}
\ee
Putting these formulas together, we can finally write (\ref{main1}) 
using only the variable $y$ \cite{JANIKPHD}
\be
R_{AB}\left(\frac{y}{R_{AB}\left(\frac{y}{R_{AB}(\ldots)}\right)}\right)  
= R_{A}\left(\frac{y}{R_{A}\left(\frac{y}{R_{A}(\ldots)}\right)}\right)
R_{B}\left(\frac{y}{R_{B}\left(\frac{y}{R_{B}(\ldots)}\right)}\right)
\ee
which amounts to the standard formulation for 
the multiplication law~\cite{VOICULESCU} as follows from (\ref{cont_frac}) 
\be
S_{AB}(y) = S_A(y) S_B(y) \ .
\label{Sab}
\ee
The necessity of assuming noncentered distributions comes from the fact 
that the implicit continued fractions appearing in (\ref{cont_frac})
make sense only for $R(z) \sim const+\oo{z}$ with nonzero constant term 
\cite{VOICULESCUPRACA}.

\subsection{Multiplication law for non-hermitian matrices}

In order to derive the multiplication law for non-hermitian matrices
we combine the two formalisms outlined in previous sections. 
First we define a $2N \times 2N$ matrix $\DD$ in analogy to (\ref{HH}) 
\be
\DD= \left( \begin{array}{cc} 0 & A \\ B & 0  \end{array}\right)_{2N \times 2N}
\ee
and then duplicate it using (\ref{defhg}) 
to obtain an extended Green's function for non-hermitian matrices.
This technique has been introduced in \cite{EGNJANJURNOW}
and used for specific ensembles \cite{EGNJANJURNOW,BUR}. In this paper
we will use it to obtain a multiplication law for arbitrary (free)
nonhermitian matrices\footnote{To remind the reader, `free' means essentially that
the probability distributions of the two ensembles are independent \emph{and} that
we take the $N\to \infty$ limit.}.

This procedure leads to a four-fold matricial structure 
("double doubling") where the primary object is a $4N \times 4N$ matrix
\be
\HH = \left( \begin{array}{cc} \DD & 0 \\ 0 & \DD^\dagger \end{array}\right) =
\left(\begin{array}{cccc} 0&A&0&0 \\ B&0&0&0 \\ 0&0&0&B^{\dagger} \\
 0&0&A^{\dagger}&0
\end{array}\right)_{4N \times 4N}
\label{HH4}
\ee
and the corresponding Green's function
\be 
\GG(w,\bar{w}) &=&
\left\langle\left[ \left(
\begin{array}{cccc} w\mathbbm{1}&0&0&0 \\ 0&w\mathbbm{1}&0&0 \\ 0&0&\bar{w}\mathbbm{1}&0
 \\ 0&0&0&\bar{w}\mathbbm{1}
\end{array}
\right)
- \left(
\begin{array}{cccc} 0&A&0&0 \\ B&0&0&0 \\ 0&0&0&B^{\dagger} \\
 0&0&A^{\dagger}&0
\end{array}\right)
\right]^{-1}\right\rangle
\label{case2green}
\ee
Using the block-trace operation $\tr_{b4}$ we reduce
the problem to calculations for $4 \times 4$ matrices
\be \GG({\cal W}) \equiv \left(
\begin{array}{cccc} \GG_{11}&\GG_{12}&\GG_{1\bar{1}}&\GG_{1\bar{2}} \\
\GG_{21}&\GG_{22}&\GG_{2\bar{1}}&\GG_{2\bar{2}} \\
\GG_{\bar{1}1}&\GG_{\bar{1}2}&\GG_{\bar{1}\bar{1}}&\GG_{\bar{1}\bar{2}} \\
\GG_{\bar{2}1}&\GG_{\bar{2}2}&\GG_{\bar{2}\bar{1}}&\GG_{\bar{2}\bar{2}} \\
\end{array}
\right)_{4 \times 4} = \frac{1}{N} \tr_{b4}{\GG}(w,\bar{w})
\label{case2green4} 
\ee 
where ${\cal W} = {\rm diag}(w,w,\bar{w},\bar{w})$.
The labeling of the matrix elements follows the convention
adopted in the previous sections. Similarly, we define 
a self-energy $\Sigma({\cal W})$ as a $4$ by $4$ matrix:
 \be \GG(W)= \left(\WW-{\Sigma}(\WW)\right)^{-1} 
 \label{DSI}
\ee  
which is related to a $4$ by $4$ matrix representing the generalized R transform:
\be
\Sigma(\WW) = \RR(\GG(\WW)) \ .
\label{DSII}
\ee
The elements of $\Sigma$ and $\RR$ are indexed in the same way as the elements
of $\GG$ (\ref{case2green4}).  

We exploit  these $4 \times 4$ matrices as auxiliary objects 
to derive relations between $2 \times 2$ Green's functions
$\GG_A(\ZZ)$, $\GG_B(\ZZ)$ and $\GG_M(\ZZ)$
for $A$, $B$ and the product $M=AB$. The naming convention 
for elements of $2\times 2$ matrices  
\be
\GG_A(\ZZ) = \left( 
\begin{array}{cc} 
\GG_{(A)11} & \GG_{(A)1\bar{1}} \\
\GG_{(A)\bar{1}1} & \GG_{(A)\bar{1}\bar{1}}
\end{array} \right)
\ee
is a bit inconvenient since it requires three subscripts for each element.
To avoid multiple subscripts like $(A)1\bar{1}$ 
we introduce a shorthand notation  substituting  multiples 
indices like $(A)1\bar{1}$ by  $A\bar{A}$ etc. In this new
notation a double subscript identifies both the matrix for which 
the generating function is calculated and the position 
of the element. Using this convention we have
\be
\GG_A(\ZZ) = \left( \begin{array}{cc} 
\GG_{AA} & \GG_{A\bar{A}} \\
\GG_{\bar{A}A} & \GG_{\bar{A}\bar{A}}
\end{array} \right)
\ee
and similarly for two remaining generating functions
\be
\Sigma_A(\ZZ) = \left( \begin{array}{cc} 
\Sigma_{AA} & \Sigma_{A\bar{A}} \\
\Sigma_{\bar{A}A} & \Sigma_{\bar{A}\bar{A}}
\end{array} \right) , \
{\cal R}_A(\GG) = \left( \begin{array}{cc} 
{\cal R}_{AA} & {\cal R}_{A\bar{A}} \\
{\cal R}_{\bar{A}A} & {\cal R}_{\bar{A}\bar{A}}
\end{array} \right) \ .
\ee
We use the same convention for all matrices, including $B$ and $M$. 
For brevity we skipped the arguments of the matrix elements 
on the right hand side of the equations above. We tacitly assumed
that they are identical as on the left hand side. 
We will frequently use this shorthand notation below.

To summarize the notation, ${\cal R}_M$ denotes a $2 \times 2$ 
matrix of the R transform for $M$ while ${\cal R}_{MM}$ -- its upper 
left element, etc. For $4 \times 4$ matrices like
${\cal G}(\WW)$, $\Sigma(\WW)$ and $\RR(\GG)$ we instead use
the indexing as in (\ref{case2green4}) which uniquely
identifies the positions of elements in such $4\times 4$ matrices. 
The link between the two conventions emerges from the equation (\ref{HH4}) 
\be
\HH \equiv 
\left(
\begin{array}{cccc} \HH_{11}&\HH_{12}&\HH_{1\bar{1}}&\HH_{1\bar{2}} \\
\HH_{21}&\HH_{22}&\HH_{2\bar{1}}&\HH_{2\bar{2}} \\
\HH_{\bar{1}1}&\HH_{\bar{1}2}&\HH_{\bar{1}\bar{1}}&\HH_{\bar{1}\bar{2}} \\
\HH_{\bar{2}1}&\HH_{\bar{2}2}&\HH_{\bar{2}\bar{1}}&\HH_{\bar{2}\bar{2}} \\
\end{array}
\right)
=
\left(\begin{array}{cccc} 0&A&0&0 \\ B&0&0&0 \\ 0&0&0&B^{\dagger} \\
 0&0&A^{\dagger}&0
\end{array}\right)
\label{HAB}
\ee
that allows us to identify $A \leftrightarrow \HH_{12}$, 
$A^\dagger \leftrightarrow \HH_{\bar{2}\bar{1}}$ and 
$B \leftrightarrow \HH_{21}$, $B^\dagger \leftrightarrow \HH_{\bar{1}\bar{2}}$.
We use this identification to rewrite the equation (\ref{DSII})
in terms of $2 \times 2$ matrices. We begin by noting that
even powers $\HH^{2k}$ of $\HH$ (\ref{HAB}) generate powers $M^k$ of 
the product $M=AB$
in the upper left corner of the block matrices
\be
\HH^{2k} =
\left(\begin{array}{cccc} (AB)^k & 0 &0&0 \\ 0& (BA)^k &0&0 \\ 0&0& (AB)^{\dagger k} & 0 \\
 0&0& 0 & (BA)^{\dagger k}
\end{array}\right)
\label{H2k}
\ee
These moments are generated by the element $\GG_{11}(\WW)$ 
of the $4 \times 4$ Green's function $\GG(\WW)$ (\ref{case2green4}) or alternatively by the element $\GG_{MM}(\ZZ)$ of the $2 \times 2$ Green's 
function $\GG_M(\ZZ)$, so we have
\be
G_{AB}(z,\bar{z}) = \GG_{MM}(\ZZ) = \frac{\GG_{11}(\WW)}{w}
\label{GGG}
\ee
where $\ZZ = {\rm diag}(z,\bar{z})$, $\WW = {\rm diag}(w,w,\bar{w},\bar{w})$
and $z=w^2$, analogously to (\ref{GabG11}). This equation, allows us to determine Green's  function 
$G_{AB}(z,\bar{z})$ and additionally  provides a link between $\GG_{M}$ and $\GG_{A}$ and $\GG_B$ since elements of the $4 \times 4$ Green's function $\GG$ can be explicitly expressed in terms of $\GG_A$ and $\GG_B$, as we will see below using planar Feynman diagrams.

\begin{figure}
\begin{center}
\includegraphics[width=9cm]{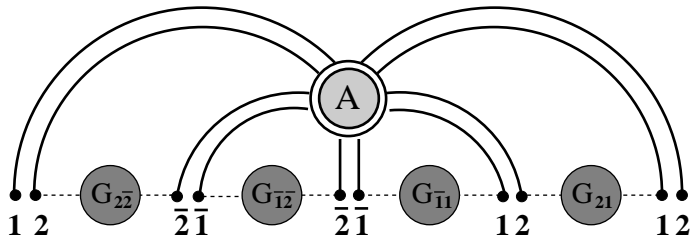} 
\caption{An example of $A$-spider connected diagram. 
Such diagrams are generated by sequences of $A$ and  $A^\dagger$
which due to the correspondence (\ref{HAB})  $A \leftrightarrow \HH_{12}$, 
$A^\dagger \leftrightarrow \HH_{\bar{2}\bar{1}}$ generate
sequences of pairs $12$ and $\bar{2}\bar{1}$. In this example
we have a sequence $12$,$\bar{2}\bar{1}$,$\bar{2}\bar{1}$,$12$,$12$
which begins with the index $1$ and ends with $2$. It contributes to $\Sigma_{12}=\RR_{12}(\GG)$, a product of
$\GG_{2\bar{2}}$, $\GG_{\bar{1}\bar{2}}$,
$\GG_{\bar{1}1}$, $\GG_{21}$ which can be read off from the
picture by matching the indices on the horizontal line. 
\label{mixed_spider}}
\end{center}
\end{figure}

First we recall that all mixed connected diagrams vanish since 
$AB$ propagators are equal zero. The last statement means that there are
no direct lines in the diagram connecting $A$ and $B$ vertices. 
All non-vanishing connected diagrams are either of $A$-type like  $\langle\langle \frac{1}{N} {\rm tr} A A^\dagger AA \ldots \rangle\rangle$ or $B$-type  like $\langle\langle \frac{1}{N} {\rm tr} B B^\dagger BB \ldots \rangle\rangle$. They are generated by alternating  
sequences either of $A$ and $A^\dagger$ or of $B$ and $B^\dagger$ but
not mixed ones.  In other words there are only $A$-spider or $B$-spider
diagrams. In the $\HH$ notation the first type is generated by 
sequences of $\HH_{12}$ and $\HH_{\bar{2}\bar{1}}$ while the second type 
of $\HH_{21}$ and $\HH_{\bar{1}\bar{2}}$ as follows from the correspondance
(\ref{HAB}). We show in figure \ref{mixed_spider} an example of a diagram
contributing to the left hand side of equation (\ref{DSII}). 
More generally, diagrams with a $A$-spider have on
the horizontal line alternating sequences like 
$\HH_{12} \HH_{\bar{2}\bar{1}} \HH_{\bar{2}\bar{1}} 
\HH_{12} \HH_{12} \ldots $ which are
sandwiched by $\GG_{2\bar{2}}$, $\GG_{\bar{1}\bar{2}}$,
$\GG_{\bar{1}1}$, $\GG_{21}$,  $\ldots$ which match the index sequence.
The left most index in the sequence of $\HH$'s
may be equal $1$ or $\bar{2}$ and the right most $2$ or $\bar{1}$ so the corresponding diagrams contribute to $\Sigma_{12}$,  $\Sigma_{1\bar{1}}$, $\Sigma_{\bar{2}2}$ or $\Sigma_{\bar{2}\bar{1}}$. 
Diagrams with a $B$-spider have sequences like
$\HH_{21} \HH_{\bar{1}\bar{2}} \ldots \HH_{21}$ etc, whose left most index 
is either $2$ or $\bar{1}$ and the right most index is either $1$ or $\bar{2}$, so the corresponding diagrams contribute to $\Sigma_{21}$, $\Sigma_{2\bar{2}}$, $\Sigma_{\bar{1}1}$ or $\Sigma_{\bar{1}\bar{2}}$. All others $\Sigma$'s must be equal zero
\be
\Sigma_{11}=\Sigma_{22}=\Sigma_{\bar{1}\bar{1}}=\Sigma_{\bar{2}\bar{2}}=0
\nonumber \\
\Sigma_{1\bar{2}}=\Sigma_{2\bar{1}}=\Sigma_{\bar{1}2}=\Sigma_{\bar{2}1}=0
\label{zeros}
\ee 
since there are no mixed $AB$-spiders. Coming back to the equations for
$\Sigma_{12}$,  $\Sigma_{1\bar{1}}$, $\Sigma_{\bar{2}2}$, $\Sigma_{\bar{2}\bar{1}}$ generated by the $A$-spider we notice 
that the indices of the $\GG$ bubbles which enter the sandwich between the spider legs have complementary  indices $\GG_{2\bar{2}}$, $\GG_{\bar{1}\bar{2}}$, $\GG_{\bar{1}1}$, $\GG_{21}$ as 
compared to those of $\Sigma$'s. The same holds for equations 
for indices of $\GG$'s and $\Sigma$'s generated by the $B$-spider.
Moreover, if we compare indices of $\Sigma$'s for $A$ spiders 
to $\GG$'s for $B$ spiders we see they are identical, 
and the same holds for $\Sigma$'s for $B$ spiders and $\GG$'s
for $A$ spiders. All these observations can be concisely 
summarized by the following equation
\be  
\Sigma 
=\left(
\begin{array}{cccc} 
0 &\Sigma_{AA}&\Sigma_{A\bar{A}}& 0 \\
\Sigma_{BB}& 0 & 0 &\Sigma_{B\bar{B}} \\
\Sigma_{\bar{B}B}& 0 & 0 &\Sigma_{\bar{B}\bar{B}} \\
0 &\Sigma_{\bar{A}A}&\Sigma_{\bar{A}\bar{A}}& 0\\
\end{array}
\right)  \, .
\label{blockSigma} 
\ee
The matrix has eight zeros which correspond to (\ref{zeros}).
The remaining eight elements can be grouped in two groups
of four elements each of which can be mapped into a $2 \times 2$ matrix.
More precisely, the matrix $\Sigma$ of dimensions $4 \times 4$ 
is expressed in terms of $2 \times 2$ generating functions $\RR$ and $\GG$ 
for $A$ and $B$: 
\be
{\Sigma}_A = 
\left(\begin{array}{cc} \Sigma_{AA} & \Sigma_{A\bar{A}} \\
\Sigma_{\bar{A}A} & \Sigma_{\bar{A}\bar{A}} \end{array} \right) =
\left(\begin{array}{cc} \RR_{AA}(\GG_B) & \RR_{A\bar{A}}(\GG_B) \\
\RR_{\bar{A}A}(\GG_B) & \RR_{\bar{A}\bar{A}}(\GG_B) \end{array} \right)
= \RR_A(\GG_B)
\ee
and
\be
{\Sigma}_B = 
\left(\begin{array}{cc} \Sigma_{BB} & \Sigma_{B\bar{B}} \\
\Sigma_{\bar{B}B} & \Sigma_{\bar{B}\bar{B}} \end{array} \right) =
\left(\begin{array}{cc} \RR_{BB}(\GG_A) & \RR_{B\bar{B}}(\GG_A) \\
\RR_{\bar{B}B}(\GG_A) & \RR_{\bar{B}\bar{B}}(\GG_A) \end{array} \right)
= \RR_B(\GG_A)
\ee
The argument of $\RR_A$ in $\Sigma_A=\RR_A(\GG_B)$ is $\GG_B$ 
while the argument of $\RR_B$ in $\Sigma_B = \RR_B(\GG_A)$ is $\GG_A$
as argued above where
\be
\GG_A =
\arr{\GG_{12}}{\GG_{1\bar{1}}}{\GG_{\bar{2}2}}{\GG_{\bar{2}\bar{1}}}
 \ , \quad
\GG_B=
\arr{\GG_{21}}{\GG_{2\bar{2}}}{\GG_{\bar{1}1}}{\GG_{\bar{1}\bar{2}}} \ .
\label{GAGB}
 \ee
So far we have used diagrammatic properties of the equation (\ref{DSII}).
Now we can also exploit the second equation (\ref{DSI}).
Inverting the matrix on the left hand side of this equation
for the particular form (\ref{blockSigma}) we can find elements
of $\GG$ as functions of $\Sigma$'s.
In particular the equation for $\GG_{11}$ is
\be
\GG_{11} = \frac{w\bar{w}^2 - \bar{w} \Sigma_{\bar{A}A} \Sigma_{B\bar{B}}
- w \Sigma_{\bar{A}\bar{A}} \Sigma_{\bar{B}\bar{B}}}{{\rm det}(\WW - \Sigma)}
\label{gg11}
\ee
where
\be
\det({\cal{W}}-\Sigma)&=&w^2\bar{w}^2
-w^2\Sigma_{\bar{A}\bar{A}}\Sigma_{\bar{B}\bar{B}}
-\bar{w}^2\Sigma_{AA}\Sigma_{BB} \nonumber \\
&+&
(\Sigma_{A\bar{A}}\Sigma_{\bar{A}A}-\Sigma_{AA}\Sigma_{\bar{A}\bar{A}})
(\Sigma_{B\bar{B}}\Sigma_{\bar{B}B}-\Sigma_{BB}\Sigma_{\bar{B}\bar{B}}) \nonumber \\
&-&\bar{w}w (\Sigma_{A\bar{A}}\Sigma_{\bar{B}B} +
\Sigma_{\bar{A}A}\Sigma_{B\bar{B}}) \label{almostfact} 
\ee 
Now we can use (\ref{GGG}) to compare $\GG_{11}/w$ that follows from
(\ref{gg11}) to $\GG_{MM}$
\be
\GG_{MM} = \frac{\bar{z} - \Sigma_{\bar{M}\bar{M}}}{{\rm det}(\ZZ - {\bf \Sigma}_M)} 
\ee
where 
\be
{\bf \Sigma}_M = 
(z-\Sigma_{MM})(\bar{z}-\Sigma_{\bar{MM}})-\Sigma_{M\bar{M}}\Sigma_{\bar{M}M}
\label{factor} 
\ee 
and $z=w^2$. From this comparison we can deduce relations between
$\Sigma_M$ and $\Sigma_A$ and $\Sigma_B$. The numerators 
of expressions for $\GG_{11}/w$ and of $\GG_{MM}$ are equal if
\be
\Sigma_{\bar{M}\bar{M}} = \frac{\bar{w}}{w} \Sigma_{\bar{A}A} \Sigma_{B\bar{B}}
+ \Sigma_{\bar{A}\bar{A}} \Sigma_{\bar{B}\bar{B}}
\ee
and the denominators (\ref{almostfact}), (\ref{factor}) if
\be
(w^2-\Sigma_{MM}
&)(&\bar{w}^2-\Sigma_{\bar{M}\bar{M}})-\Sigma_{M\bar{M}}\Sigma_{\bar{M}M} \nonumber \\ &=& w^2\bar{w}^2 -w^2\Sigma_{\bar{A}\bar{A}}\Sigma_{\bar{B}\bar{B}}
  - \bar{w}^2\Sigma_{AA}\Sigma_{BB} \nonumber \\ 
&+& (\Sigma_{A\bar{A}}\Sigma_{\bar{A}A}-\Sigma_{AA}\Sigma_{\bar{A}\bar{A}})
(\Sigma_{B\bar{B}}\Sigma_{\bar{B}B}-\Sigma_{BB}\Sigma_{\bar{B}\bar{B}}) \\
&-& \bar{w}w (\Sigma_{A\bar{A}}\Sigma_{\bar{B}B} +
\Sigma_{\bar{A}A}\Sigma_{B\bar{B}}) \, . \nonumber
\ee
One can check that the two equations are simultaneously fulfilled if
\be
\Sigma_{MM} &=& \Sigma_{AA}\Sigma_{BB} + 
\frac{w}{\bar{w}} \Sigma_{A\bar{A}}\Sigma_{\bar{B}B} \nonumber \\
\Sigma_{M\bar{M}} &=& \sqrt{\frac{\bar{w}}{w}} \Sigma_{AA}\Sigma_{B\bar{B}} +
\sqrt{\frac{w}{\bar{w}}} \Sigma_{A\bar{A}}\Sigma_{\bar{B}\bar{B}} \nonumber \\
\Sigma_{\bar{M}M} &=& \sqrt{\frac{\bar{w}}{w}} \Sigma_{\bar{A}A}\Sigma_{BB} +
\sqrt{\frac{w}{\bar{w}}} \Sigma_{\bar{A}\bar{A}}\Sigma_{\bar{B}B} \\
\Sigma_{\bar{M}\bar{M}} &=& \Sigma_{\bar{A}\bar{A}} \Sigma_{\bar{B}\bar{B}} +
\frac{\bar{w}}{w} \Sigma_{\bar{A}A} \Sigma_{B\bar{B}} \, .\nonumber
\ee
Remarkably, these equalities can be written in a {\it factorizable} matrix form as
\be
 {\Sigma}_M
&\equiv& \setlength\arraycolsep{3pt}
\arr{\Sigma_{MM}}{\Sigma_{M\bar{M}}}{\Sigma_{\bar{M}M}}{\Sigma_{\bar{M}\bar{M}}}
= \setlength\arraycolsep{3pt}
\arr{\Sigma_{AA}}{\sqrt{\frac{w}{\bar{w}}}\Sigma_{A\bar{A}}}{\sqrt{\frac{\bar{w}}{w}}\Sigma_{\bar{A}A}}{\Sigma_{\bar{A}\bar{A}}}
\cdot 
\setlength\arraycolsep{3pt}
\arr{\Sigma_{BB}}{\sqrt{\frac{\bar{w}}{w}}\Sigma_{B\bar{B}}}{\sqrt{\frac{w}{\bar{w}}}\Sigma_{\bar{B}B}}{\Sigma_{\bar{B}\bar{B}}}\nonumber\\
&\equiv& {\Sigma}_A^L {\Sigma}_B^R \ .
\label{SSS}
\ee
In order to simplify the notation it is convenient to introduce
a $2\times 2$ unitary diagonal matrix $U$ 
\be U &\equiv& \setlength\arraycolsep{3pt}
\arr{\left(\frac{w}{\bar{w}}\right)^{1/4}}{0}{0}{\left(
\frac{\bar{w}}{w}\right)^{1/4}} = 
\arr{e^{+i\frac{\psi}{2}}}{0}{0}{e^{-i\frac{\psi}{2}}}\eqnx 
where the angle $\psi$ is the phase of $w$: $w= |w|e^{i\psi}$. Note that $w$ is related to the original variable $z$ as 
$z=w^2$, so Arg $z$ = 2 Arg $w$. 
Using this matrix we can associate with any matrix $X$
two similar matrices $X^L$ and $X^R$ 
obtained by "left and right $U$-rotations" of the matrix in question
\be 
X^L \equiv [X]^L=U X U^{\dagger} \ , \quad 
X^R \equiv [X]^R=U^{\dagger}X U \ ,
\label{LRrot}
\ee 
In particular 
\be 
\Sigma_A^L\equiv\left[\Sigma_A\right]^L=U\Sigma_AU^{\dagger} \ , \quad
\Sigma_A^R\equiv\left[\Sigma_B\right]^R=U^{\dagger}\Sigma_B U \ .
\ee
The operations $[\ldots]^L$ and $[\ldots]^R$ obey simple rules
like for instance
\be
[XY]^L = [X]^L [Y]^L = X^LY^L , \, \left[X^{-1}\right]^L = \left([X]^L\right)^{-1} , \, X= \left[[X]^L\right]^R \ .
\ee
which we will frequently use below.

Now we come to the main result of the paper. 
Recalling that $\Sigma_A = \RR_A(\GG_B)$ and $\Sigma_A = \RR_A(\GG_B)$ we have (\ref{SSS})  
\be
\RR_M(\GG_M) = \left[\RR_A(\GG_B)\right]^L \cdot \left[\RR_B(\GG_A)\right]^R \ .
\label{FINAL1}
\ee
This equation is a cornerstone of the matrix multiplication 
for non-hermitian matrices. Let us note the similarity with the corresponding
equation for the hermitian case (\ref{main1}) albeit with two key differences.
Firstly, the objects appearing in (\ref{FINAL1}) are generically noncommuting
$2\times 2$ matrices and hence the ordering is crucial. Secondly, the left- and
right- $U$-rotations have no analogue in the scalar hermitian case. 

In fact, to arrive to
this point we have only taken advantage of the equations
for the element $\GG_{11}$ of the $4\times 4$ Green's function.
Inverting the matrix on the right hand side of (\ref{DSI})  
for $\Sigma$ given by (\ref{blockSigma})
we can relate remaining elements of $\GG$ to the elements of 
$2\times 2$ $\Sigma$'s and $\RR$'s. 
In particular we can write equations for elements
$\GG_{12}$, $\GG_{1\bar{1}}$, $\GG_{\bar{2}2}$ and $\GG_{\bar{2}\bar{1}}$
which, as we know (\ref{GAGB}) form a $2\times 2$ matrix corresponding to
the Green's function $\GG_A$ and similarly for $\GG_{21}$, $\GG_{2\bar{2}}$, $\GG_{\bar{1}1}$ and $\GG_{\bar{1}\bar{2}}$ corresponding to $\GG_B$. 
This allows us to express $\GG_A$ and $\GG_B$ in terms of $\Sigma_A$ and $\Sigma_B$. After some straightforward but tedious algebra we
arrive at remarkably simple equations which again are analogs of the
hermitian equations (\ref{main2}) but with specific ordering and appropriate 
$U$-rotations
\be
\GG_{A} \!&\!=\!&\! 
\left[\GG_M \cdot \left[\RR_A(\GG_B)\right]^L\right]^L  \nonumber \\
\GG_{B} \!&\!=\!&\! \left[\left[\RR_B(\GG_A)\right]^R \cdot \GG_M\right]^R  \ .
\label{FINAL2}
\ee
The set of equations (\ref{FINAL1}) and (\ref{FINAL2}) gives 
the multiplication law for non-hermitian matrices and constitutes 
one of main results of this work, as mentioned at the beginning
of the paper (\ref{AB_G}). 

These equations are in one-to one-correspondence 
to (\ref{main1}) and (\ref{main2}) except that now instead of complex
numbers $g_M$, $g_A$ and $g_B$ we have $2\times 2$ matrices $\GG_M$,
$\GG_A$ and $\GG_B$ and the additional $U$-rotations. 
The logic of the method to calculate
the Green's function for the product $M=AB$ is the same as for
hermitian matrices, that is for given $\GG_A$ and $\GG_B$ one
determines the matricial R transforms $\RR_A$ and $\RR_B$ and
then applies (\ref{FINAL1}) and (\ref{FINAL2}) to derive
the Green's function for $M$. We will present  examples 
in the next section. Before doing that let us  show how 
these equations can be reformulated in terms of 
a nonhermitian generalization of the S transform.

\subsection{S transform for non-hermitian matrices}

It is natural to anticipate  that   the S transform for non-hermitian matrices  has a form of a $2 \times 2$ matrix.  
It will however  appear in two different "left" and "right" versions, since $2 \times 2 $ matrices do not commute in general.   
To demonstrate this, we  repeat the arguments which have guided  us from
(\ref{main1}) and (\ref{main2}) to (\ref{Sab}), but now we adapt the reasoning   
to the case of $2\times 2$ matrix-valued transforms.

The first step is to eliminate $\GG_B$ and $\GG_A$ from
the right hand side of (\ref{FINAL1}) and substitute them 
by $\GG_M$ in order to have the same argument on both sides 
of the equation. To make the following equations 
slightly more readable we shall skip the subscript $M$ of $\GG_M$ writing $\GG\equiv \GG_M$ and we will denote the inverse matrix of a matrix $X$ 
as $\frac{1}{X}$ rather than $X^{-1}$ to avoid too many superscripts.
Using (\ref{FINAL2}) and (\ref{FINAL1}) we have
\be
\GG_B = \left[\RR^R_B(\GG_A) \GG\right]^R =
\left[\frac{1}{\RR^L_A(\GG_B)} \RR_M(\GG) \GG\right]^R \ .
\ee
This is an equation for $\GG_B$ but $\GG_B$ is also present on the right hand side.
We can however eliminate $\GG_B$  by replacing 
it recursively by the right hand side and repeating this 
infinitely many times. In this way we obtain
a nested expression (denoted below by dots) 
\be
\GG_B = \left[\frac{1}{\RR^L_A(\ldots)} \RR_M(\GG) \GG \right]^R \ .
\ee
that depends on $\GG$ and not on $\GG_B$. 
Thus we can write the first factor, $\RR^L_A(\GG_B)$, on 
the right hand side of (\ref{FINAL1}) as a function of $\GG$:
\be
\RR^L_A(\GG_B) = \RR^L_A\left(   
\left[\frac{1}{\RR^L_A(\ldots)} \RR_M(\GG) \GG \right]^R
\right) = \frac{1}{\SS^{(L)}_A\left(\RR_M(\GG)\GG\right)}
\ee
where $\SS^{(L)}$ is a left S transform defined as
\be
\SS^{(L)}(\YY) =
\frac{1}{\RR^L\left(\left[\frac{1}{\RR^L(\ldots)} \YY \right]^R
\right)} \ .
\ee
Let us make two further remarks concerning the notation. In the last equation
we skipped the subscript $A$ of $\SS$ and $\RR$ since the relation is valid for
any matrix. The superscript $(L)$ of $\SS$ is used on purpose in parentheses to distinguish it from $L$ and to emphasize that the left S transform is not a left rotation 
of the S transform $\SS^{(L)} \ne [\SS]^L \equiv U S U^\dagger$ in contrast
to the notation $\RR^L = [\RR]^L$. The function $S^{(L)}$ is just 
defined by the equation above. This equation is equivalent to
\be
\SS^{(L)}(\YY) =
\frac{1}{\RR^L\left(\left[ \SS^{(L)}(\YY) \YY \right]^R 
\right)} 
\ee
and 
\be
\RR^{L}(\YY) =
\frac{1}{\SS^{(L)}\left(\left[\RR(\YY) \YY\right]^L \right)} \ .
\label{rls}
\ee
in analogy to the hermitian case discussed in section \ref{SRrelations}.
Now we can repeat all the steps for the second factor on
the right hand side of (\ref{FINAL1}). The result can be written
using a right S transform, which is given by two equivalent, reciprocal, equations analogous to those of the left S transform above:
\be
\SS^{(R)}(\YY) =
\frac{1}{\RR^R\left(\left[\YY \SS^{(R)}(\YY) \right]^L 
\right)} 
\ee
or
\be
\RR^{R}(\YY) =
\frac{1}{\SS^{(R)}\left(\left[\YY \RR(\YY)\right]^R \right)} \ .
\label{rrs}
\ee
Using the left and right S transforms we can write (\ref{FINAL1}) 
in a concise form
\be
\frac{1}{\RR_{M}(\GG)} =  
\SS^{(R)}_B\left(\GG \RR_M(\GG)\right) \cdot 
\SS^{(L)}_A\left(\RR_M(\GG) \GG\right)
\label{R_SS}
\ee
that depends on $\GG$ on both sides. 
This is an equation for the R transform $\RR_M(\GG)$ which 
in turn determines the generalized Green's function giving the
eigenvalue density. 

Let us rewrite now the left-hand side using either equation (\ref{rls})
or (\ref{rrs})
\eq
\left[ \SS^{(L)}_M\left(\left[\RR(\GG) \GG\right]^L \right) \right]^R
=
\left[ \SS^{(R)}_M\left(\left[\GG \RR(\GG)\right]^R \right) \right]^L
=
\SS^{(R)}_B\left(\GG \RR_M(\GG)\right) \cdot 
\SS^{(L)}_A\left(\RR_M(\GG) \GG\right)
\eqx 
which now (almost) takes the form of a multiplication law for $\SS$ transforms
with the only subtlety being the noncommutativity of the arguments.

In the special case when $\GG$ and $\RR(\GG)$
commute
\be
[\GG, \RR_M(\GG)] = 0 
\ee
we get a direct analogue of the hermitian multiplication law for 
S transforms since all functions are evaluated on the same argument
$\YY=\GG \RR(\GG)=\RR(\GG) \GG$. In this case
it would make sense to introduce yet another S transform:
\be
\RR(\YY) =
\frac{1}{\SS\left(\RR(\YY) \YY \right)} \ ,
\ee
which does not involve any left or right $U$-rotation.
It is easy to see that in this case the equation 
(\ref{R_SS}) can be rewritten as
\be
\SS(\YY) = \SS^{(R)}_B\left(\YY\right) \cdot 
\SS^{(L)}_A\left(\YY\right) \ .
\ee
One should note that the $2\times 2$ formalism, which has been 
developed here for non-hermitian random matrix ensembles, contains also the standard hermitian 
case. For hermitian matrices, namely, the Green's functions and the R transforms reduce to diagonal matrices
\be
\GG(\ZZ) = \left(\begin{array}{cc} G(z) & 0 \\ 0 & \bar{G}(z) \end{array}\right) \ ,
\quad 
\RR(\GG) = \left(\begin{array}{cc} R(G) & 0 \\ 0 & \bar{R}(G) \end{array}\right) \ ,
\ee
Moreover $\GG=\GG^L=\GG^R$, $\RR=\RR^L=\RR^R$ because the matrix 
$U$ that defines the left and right rotations (\ref{LRrot}) is diagonal too and the product of the diagonal elements gives one. It follows also that $\SS=\SS^{(R)}=\SS^{(L)}$ and that the S transform is diagonal 
$\SS(\ZZ) = {\rm diag}(S(z),\bar{S}(z))$ too. Therefore in this case
(\ref{R_SS}) takes a diagonal from
\be
\left( \begin{array}{cc} S_M(z) & 0 \\ 0 & \bar{S}_M(z) \end{array}\right) =
\left( \begin{array}{cc} S_A(z) S_B(z) & 0 \\ 0 & \bar{S}_A(z) \bar{S}_B(z) \end{array}\right)
\ee
that is equivalent to (\ref{Sab}).

\section{Examples \label{Sec_examples}}

In this section we will illustrate our methods by presenting three examples. We will
start from two examples which cannot be treated, even in the hermitian case,
by the conventional S transform treatment as \emph{both} of the random matrix 
factors of the product are centered. Finally we treat a more complicated example
of obtaining a nontrivial two-dimensional eigenvalue distribution for a product
of two simple factors.

\subsection{Product of two free Ginibre-Girko matrices}
 
Let us first consider the product $M=AB$ of two identically distributed
free Ginibre-Girko matrices: $A$ and $B$. Throughout this section
we will parametrize matrix elements of the $2 \times 2$ Green's functions (\ref{19}) with two complex functions
$a=a(z,\bar{z})$ and $b=b(z,\bar{z})$
\be 
\GG = 
\arr{{\GG}_{11}}{{\GG}_{1\bar{1}}}{{\GG}_{\bar{1}1}}
{{\GG}_{\bar{1}\bar{1}}} = \left(
\begin{array}{cc} a & ib \\ i\bar{b} & \bar{a} \end{array} \right)
\ee
The R transform for a 
Ginibre-Girko matrix reads (\ref{gin2}).
\be
\RR(\GG)= \RR\left(\left(
 \begin{array}{cc}
  a & ib \\ i\bar{b} & \bar{a} \end{array} \right)\right) =
   \left(\begin{array}{cc}
  0 & ib \\ i\bar{b} & 0 \end{array} \right) 
\label{RGG}
\ee
and its left and right versions
\be
\RR^L(\GG)  = \left(\begin{array}{cc}
  0 & \sqrt{\frac{\bar{w}}{w}} i b  \\  \sqrt{\frac{w}{\bar{w}}} i \bar{b}  & 0 \end{array} \right) \ , \quad
  \RR^R(\GG)  = \left(\begin{array}{cc}
  0 & \sqrt{\frac{w}{\bar{w}}} i b  \\ \sqrt{\frac{\bar{w}}{w}} i \bar{b}  & 0 \end{array} \right) 
\ee
respectively. We recall that $w$ is related to $z$ as $z=w^2$. Let us now
apply the multiplication law for $M=AB$ where $A$ and $B$ are Ginibre-Girko
matrices with unit variance. Using (\ref{FINAL1}) we have
\be
\RR_M =  
 \left(\begin{array}{cc}
  0 & \sqrt{\frac{w}{\bar{w}}} i b_B  \\ \sqrt{\frac{\bar{w}}{w}} i \bar{b}_B  & 0 \end{array} \right) 
 \left(\begin{array}{cc}
  0 & \sqrt{\frac{\bar{w}}{w}} i b_A  \\  \sqrt{\frac{w}{\bar{w}}} i \bar{b}_A  & 0 \end{array} \right) = 
\left(\begin{array}{cc}
  - \frac{w}{\bar{w}} b_B \bar{b}_A & 0 \\ 0 & - \frac{\bar{w}}{w} \bar{b}_B b_A  \end{array} \right) 
\ee
Since both $A$ and $B$ are identically distributed they have identical
Green's function, we can thus reduce the problem by introducing 
a single function $b=b_A=b_B$:
\be
 \RR_M= \left(\begin{array}{cc}
  - \frac{w}{\bar{w}} |b|^2  & 0 \\ 0 & - \frac{\bar{w}}{w} |b|^2 
  \end{array} \right) 
  \label{RM}
\ee
We can now use the two remaining equations of the multiplication
law (\ref{FINAL2}) which can be conveniently written as
\eqn
\GG_M^{-1} \left[\GG_A\right]^R &=& \left[\RR_A(\GG_B)\right]^L \nonumber\\
\left[\GG_B\right]^L \GG_M^{-1} &=& \left[\RR_B(\GG_A)\right]^R
\label{niceform}
\eqnx
In case of identically distributed $A$ and $B$ one of the two equations 
is redundant and thus it is sufficient to use only one of them, for instance
the first one. We first eliminate $\GG_M$ from this equation
by using the relation $\GG_M^{-1}=\ZZ-\RR_M$ with $\RR_M$ given by (\ref{RM}):
\be
(\ZZ-\RR_M) \left[\GG_A\right]^R = \left[\RR_A(\GG_B)\right]^L \ .
\label{ZRG}
\ee
This is an explicit equation for $a$ and $b$
\be
 \left(\begin{array}{cc}
  w^2 + \frac{w}{\bar{w}} |b|^2  & 0 \\ 0 & 
  \bar{w}^2 + \frac{\bar{w}}{w} |b|^2 
  \end{array} \right) \left(\begin{array}{cc}
  a & \sqrt{\frac{\bar{w}}{w}} i b  \\  \sqrt{\frac{w}{\bar{w}}} i \bar{b}  & \bar{a} \end{array} \right)  = \left(\begin{array}{cc}
  0 & \sqrt{\frac{w}{\bar{w}}} i b  \\ \sqrt{\frac{\bar{w}}{w}} i \bar{b}  & 0 \end{array} \right) 
\ee
It can be easily solved. It has two solutions: a trivial and $a=b=0$ 
and a non-trivial one $a=0$,  $|b|^2 = 1 - w\bar{w}$. The latter
one is equivalent to $a=0$ and $|b|^2 = 1 - \sqrt{z\bar{z}}$ when expressed in the variable $z=w^2$. This solution holds inside the unit circle: $z\bar{z}\le 1$ on the $z$ complex plane while the trivial one outside. 
The boundary of the eigenvalue distribution in the $z$ plane is given by the condition $b=0$ for the non-trivial solution which leads to
the unit circle. Inserting these solutions to (\ref{RM}) and calculating $\GG_M$ we find
\be
\GG_M(z,\zb)= \left( \begin{array}{cc} \sqrt{\frac{\zb}{z}} & 0 \\ 
0 & \sqrt{\frac{z}{\zb}} \end{array} \right) \ ,
\quad  \mbox{for} \quad |z| \le 1
\ee
or 
\be
\GG_M(z,\zb)= \left( \begin{array}{cc} z^{-1} & 0 \\ 
0 & {\zb}^{-1} \end{array} \right) \ , 
\quad  \mbox{for} \quad |z| \ge 1
\ee
from which we obtain a rotationally symmetric eigenvalue density for $|z|<1$:
\be
\rho(x,y)=\frac{1}{\pi}\frac{\partial}{\partial \zb} G(z,\zb)= \frac{1}{2\pi} \frac{1}{|z|}
\label{rhoxy}
\ee
inside the unit circle and $\rho(x,y)=0$ outside. We remind the reader that $G(z,\zb)$
is equal to the upper left element $\GG_{MM}$ of $\GG_M$.

\subsection{Product of two free GUE matrices}

We would like to discuss a simple but very interesting case
of the product $M=AB$ of two matrices from the Gaussian Unitary Ensembles. 
Both $A$ and $B$ are hermitian but their product is not.  
Since both matrices have a vanishing mean the traditional use of 
S transform leads to contradiction, as shown in~\cite{SPERAJ}. 
However, our algorithm works in this case without any problems. 

Before we apply the full non-hermitian version of the multiplication law
let us check what happens if one applies its hermitian version given by
equations (\ref{main1}) and (\ref{main2}). One can do this since
$A$ and $B$ are hermitian. However the result for $G_M(z)$ can be
interpreted only as a moments' generating function but not as
a full Green's function. In particular one cannot use it to reconstruct the eigenvalue density (\ref{recon}) since the eigenvalues are not constrained to the real axis. 

For a standardised GUE matrix we have $R(z)=z$ and thus the
multiplication law (\ref{main1}) and (\ref{main2}) simplifies to
\eq
R_{M}(g)=g_B g_A, \quad\quad g_A=g g_B, \quad\quad g_B=g g_A
\eqx
The two latter relations yield an equation $g_A= g^2 g_A$.
Its solution is $g_A=0$ giving $R_{M}(g)=0$ and hence
\eq
G_M(z)= \f{1}{z}
\eqx
The moments $m_k$ are given by coefficients at $1/z^{k+1}$ 
of the $1/z$-expansion of $G_M(z)$. We see that all they vanish
except the trivial one $m_0 = \frac{1}{N} \langle \mbox{tr} M^0 \rangle = 1$. 
Of course it does not mean that all eigenvalues of $M$ vanish.
In order to determine the eigenvalue density of $M$ one has to apply the full
multiplication law in the domain of non-hermitian matrices (\ref{FINAL1}) and
(\ref{FINAL2}). The calculation goes along the same lines as in the previous
example except that now instead of (\ref{RGG}) the R transform is 
\be
\RR(\GG)= \RR\left(\left(
 \begin{array}{cc}
  a & ib \\ i\bar{b} & \bar{a} \end{array} \right)\right) =
   \left(\begin{array}{cc}
  a & ib \\ i\bar{b} & \bar{a} \end{array} \right)  = \GG 
\ee
as follows from (\ref{general}) for $\tau=1$. It is easy to see that
the solution is exactly the same as in the previous example since for $a=0$
(which was a solution) all equations reduce to those for the previous case. This result is in agreement with the recent works~\cite{GIRKOVLAD,BUR,LIVAN}. Actually one can 
see that the same holds for any elliptic ensemble with
\be
\RR(\GG)= \RR\left(\left(
 \begin{array}{cc}
  \tau a & ib \\ i\bar{b} & \tau \bar{a} \end{array} \right)\right) =
   \left(\begin{array}{cc}
  \tau a & ib \\ i\bar{b} & \tau \bar{a} \end{array} \right)  = \GG 
\ee
since again for $a=0$ the equations are identical as before. Again
this is in agreement with~\cite{BUR} where it was shown that
even for $A$ and $B$ being from different elliptic ensembles
($\tau_A \ne \tau_B$ or $\tau_A=\tau_B$) one obtains 
the same circular law (\ref{rhoxy}).

\subsection{Pascal lima\c{c}on}
\begin{figure}
\begin{center}
\includegraphics[width=8cm]{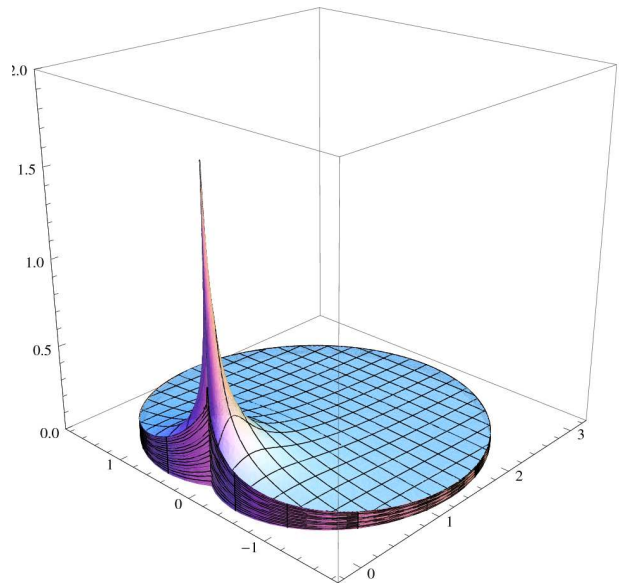}
\end{center}
\caption{The eigenvalue density of the product of two shifted Ginibre-Girko matrices. It is non-zero in the region $r \le 1 + 2\cos \phi$.
The density is peeked around the origin. The maximum 
of the function is located at the origin: 
$\rho(x=0,y=0)=\frac{6}{\pi}\approx 1.90986$ while
the minimum at the point $x=3,y=0$: 
$\rho(x=3,y=0)=\frac{9\pi}{56} \approx 0.0511569$. 
\label{density_limason}}
\end{figure}

\begin{figure}
\includegraphics[width=6cm]{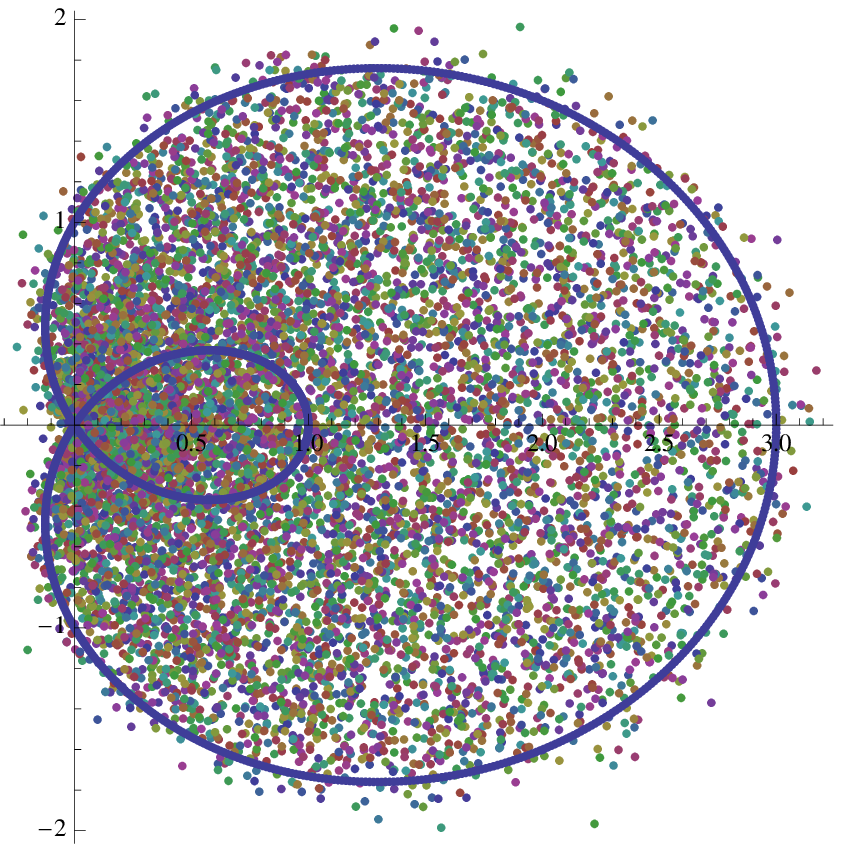}
\includegraphics[width=8cm]{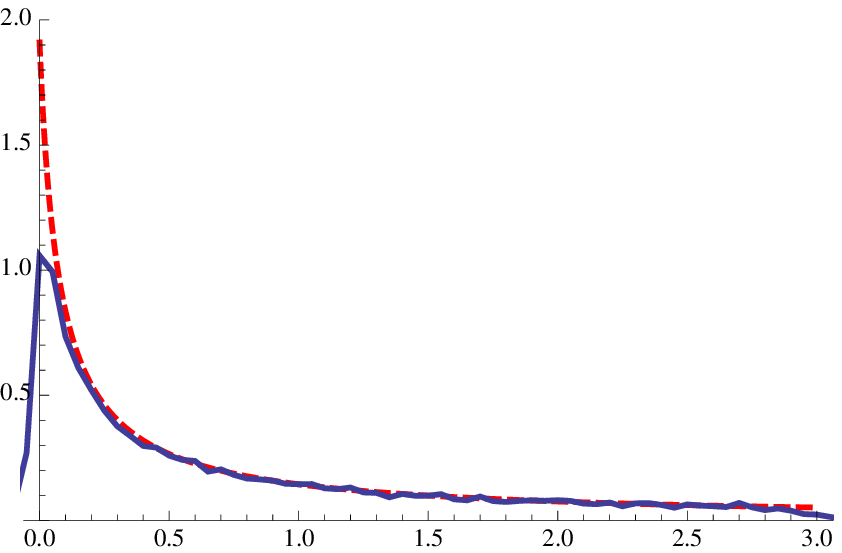}
\caption{(Left) The analytical contour $r=1+2\cos\phi$ (\ref{PasLim}) and 
the scattered plot of eigenvalues obtained by diagonalisation of $100$ 
matrices of dimensions $100\times 100$. 
One should note that the boundary of the support is formed only
by the external part of the Pascal's lima\c{c}on which corresponds to 
$\phi \in [-2\pi/3,2\pi/3]$. The remaining part of the lima\c{c}on 
lies inside the support.  (Right) A numerical histogram (solid line) constructed 
from almost real eigenvalues (whose imaginary part is less than $\epsilon=10^{-2}$) obtained by diagonalisation of $20000$ 
matrices of size $100 \times 100$ compared to the section of the analytic eigenvalue density $\rho$ along the real axis. The deviations between
the numerical histogram and the theoretical curve are caused by
finite size effects. 
\label{comparison}}  
\end{figure}

We shall calculate now the eigenvalue distribution of the product of 
two shifted Ginibre-Girko matrices $M=AB = (1+X_A)(1+X_B)$ 
where $X_A$ and $X_B$ are free Ginibre-Girko complex matrices. 
The main difference to the cases discussed before is that the multiplied
matrices $A$ and $B$ are not centered: $\frac{1}{N} \mbox{tr} A = 1$ and 
$\frac{1}{N} \mbox{tr} B = 1$, so their first moments (cumulant) are not zero:
\eqn\setlength\arraycolsep{3pt}
\RR_A(\GG_B)&=&\RR_A \left(\arr{a_B}{ib_B}{i\bar{b}_B}
        {\bar{a}_B}\right)=
\arr{1}{ib_B}{i\bar{b}_B}{1} \nonumber \\
\RR_B(\GG_A)&=&\RR_B \left(\arr{a_A}{ib_A}{i\bar{b}_A}
        {\bar{a}_A}\right)=
\arr{1}{ib_A}{i\bar{b}_A}{1} \label{product2GG} \ee 
Since $A$ and $B$ are identically distributed we set $b=b_A=b_B$ as in
the previous examples. Using (\ref{FINAL1}) we have
\be
 \RR_M= \left(\begin{array}{cc}
   1 - \frac{w}{\bar{w}}|b|^2  & ib \left(\sqrt{\frac{\bar{w}}{w}} + \sqrt{\frac{w}{\bar{w}}}\right) \\ i\bar{b}
  \left(\sqrt{\frac{\bar{w}}{w}} + \sqrt{\frac{w}{\bar{w}}}\right)
  & 1 - \frac{\bar{w}}{w} |b|^2 
  \end{array} \right) 
  \label{RM2}
\ee
Inserting this to (\ref{ZRG}) we obtain an explicit equation
\be
 \left(\begin{array}{cc}
  w^2 -1 + \frac{w}{\bar{w}} |b|^2  & -ib \left(\sqrt{\frac{\bar{w}}{w}} + \sqrt{\frac{w}{\bar{w}}}\right) \\ -i\bar{b}
  \left(\sqrt{\frac{\bar{w}}{w}} + \sqrt{\frac{w}{\bar{w}}}\right) & 
  \bar{w}^2 - 1 + \frac{\bar{w}}{w} |b|^2 
  \end{array} \right) \left(\begin{array}{cc}
  a & \sqrt{\frac{\bar{w}}{w}} i b  \\  \sqrt{\frac{w}{\bar{w}}} i \bar{b}  & \bar{a} \end{array} \right)  = \left(\begin{array}{cc}
  1 & \sqrt{\frac{w}{\bar{w}}} i b  \\ \sqrt{\frac{\bar{w}}{w}} i \bar{b}  & 1 \end{array} \right) 
\ee
which reduces to two equations for $a$ and $|b|^2$:
\be a(w^2-1+|b|^2\frac{w}{\bar{w}}) + |b|^2(1+
\frac{w}{\bar{w}})&=&1 \nonumber \\
-a(1+\frac{w}{\bar{w}})+ \frac{w}{\bar{w}}(\bar{w}^2-1)+|b|^2&=&1 \ee
This set of equations has a trivial solution: $b=0$ and $a=1/(w^2-1)$
and a non-trivial one that can be found by eliminating $a$ from the last set of equations. This gives an equation for $C=|b|^2$ (\ref{Cdef}):
\be
C^2 + C (1+2|w|^2) + |w|^4 - |w|^2 - \bar{w}^2 - w^2 = 0
\label{b4}
\ee
The border line between the two solutions can be found by setting $C=0$
in the last equation \cite{EGNJANJURNOW}:
\be \bar{w}^2w^2-\bar{w}w =w^2+\bar{w}^2 \label{limacon}
\ee 
It represents a curve on the $z$-plane called Pascal's lima\c{c}on after Etienne Pascal (1588-1651) - the father of Blaise Pascal.
It has a more familiar form in polar coordinates  on the $z$-plane:
$w^2 \equiv z=r \exp i \phi$:
\be
r=1+2 \cos \phi .
\label{PasLim}
\ee 
It is a particular case of the trisectrix. The trivial solution holds outside the Pascal's lima\c{c}on while the non-trivial inside. For the trivial solution the Green's function is $G=\GG_{MM}=1/(z-1)$ and thus $\rho(x,y)=0$. The non-trivial solution can be found by inverting $\GG_M = (\ZZ - \RR_M)^{-1}$ for $\RR_M$ (\ref{RM2}). The Green's function is given by the upper left element of $\GG_M$ \be
G=\GG_{MM} = \frac{\bar{w}^2-1+\frac{\bar{w}}{w}C}
{\left(w^2-1+\frac{w}{\bar{w}}C)(\bar{w}^2-1+\frac{\bar{w}}{w}C\right) +
C \left(\frac{\bar{w}}{w} + 2 + \frac{w}{\bar{w}}\right)}
\label{G4}
\ee
with $C$ being a solution of (\ref{b4}), and again agrees with~\cite{EGNJANJURNOW}.
 One can write down the solution
in polar coordinates on the $z$-plane: $z=re^{i\phi}$
as
\be
G \equiv G_x - iG_y = \frac{(r+C)\cos \phi - 1}{D}  - i \frac{(r+C) \sin \phi }{D} 
\ee
where $C$ and $D$ are real non-negative functions
\be
C = \frac{1}{2}\left(-1-2r + \sqrt{1 + 8r(1+\cos \phi)}\right)
\ee
and
\be
D=\left((r+C)\cos \phi - 1\right)^2 + (r+C)^2 \sin^2 \phi + 2C(1+\cos \phi)
\ee
The first one corresponds to (\ref{b4}) and the second one
to the denominator in (\ref{G4}). One can explicitly see that $C$ is positive 
inside the Pascal  lima\c{c}on $r< 1 + 2\cos \phi$. Using the Gauss law
we find the eigenvalue density
\be
\rho = \frac{1}{\pi} \frac{\partial G}{\partial \bar{z}} =
\frac{1}{2\pi} \left( \frac{\partial G_x}{\partial x} +
 \frac{\partial G_y}{\partial y} \right) = \frac{1}{2\pi} \mbox{div} \, \vec{G} \ .
\ee
The imaginary part of $\partial_{\bar{z}} G$ is proportional to
the rotation $\mbox{rot} \, \vec{G} = \partial_x G_y - \partial_y G_x$
that vanishes by construction. This fact can be used as a test of correctness
of calculations. The density calculated from this formula is shown in
figure \ref{density_limason}.
Finally, we perform some numerical checks. We generate numerically
matrices $M=AB=(1+X_A)(1+X_B)$ of dimensions 
$100\times 100$ and compare obtained eigenvalue histograms 
with the exact solution for infinite dimensions. In figure \ref{comparison} 
we show a scattered plot of eigenvalues and the histogram of 
real eigenvalues compared to the section of the analytic solution 
along the real axis. The results show a good agreement between numerical
data and the analytic result. The small remaining deviations can be attributed to finite size effects.

\section{Summary \label{Sec_summary}}

We have introduced a natural generalization of the  concept of S transform for the product of non-hermitian ensembles.
This construction puts on the same footing addition and multiplication laws for  hermitian and non-hermitian ensembles. We have also found a more general reformulation
of the multiplication law which allows us to calculate free products of
random matrices having vanishing mean, including the case when both factors
in the product are centered. This case is especially interesting as it cannot be addressed using ordinary S transform techniques.

Our construction relies on the insights from diagrammatic techniques, and in particular 
assumes the finiteness of the moments. We are however convinced, that these conditions 
are neither restrictive nor mandatory for a general proof, based on purely algebraic 
structures like e.g. amalgamation of free random variables and a careful treatment
of regularization of ensembles with ubounded moments.

\subsubsection*{Acknowledgments}
This work was partially supported by the Polish Ministry of Science
 Grants No. N N202 22913 and N N202 105136.  ZB and MAN would like to thank
the Nordita at Stockholm, where a part of this work has been
completed, for the hospitality. Their stay at Nordita was supported
by the program "Random Geometry and Applications".

\end{document}